\documentclass[aps,twocolumn,showpacs,nofootinbib]{revtex4-1}

\usepackage{color} 
\definecolor{green}{rgb}{0,.5,0}
\definecolor{red}{rgb}{1,0,0}
\usepackage{graphicx}
\usepackage[subfigure]{graphfig}
\usepackage{epsfig}
\usepackage{slashed}
\usepackage{dcolumn}
\usepackage{amsmath}
\usepackage{latexsym}
\usepackage{multirow}

\newcommand{\QCD}{\text{QCD}}
\newcommand{\GeV}{\text{GeV}}

\newcommand{\MOM}{\text{RI/MOM}}
\newcommand{\RIp}{\text{RI}^\prime}
\newcommand{\RIpMOM}{\text{RI}^\prime/\text{MOM}}
\newcommand{\SMOM}{\text{RI/SMOM}}
\newcommand{\IMOM}{\text{IMOM}}
\newcommand{\MSbar}{{\overline{\text{MS}}}}

\DeclareMathOperator{\Tr}{Tr}

\def\be{\begin{equation}}
\def\ee{\end{equation}}
\def\bea{\begin{eqnarray}}
\def\eea{\end{eqnarray}}
\def\non{\nonumber}

\begin{document}

\title{RI/MOM and RI/SMOM renormalization of quark bilinear operators using overlap fermions}

\author{Fangcheng He$^{1}$, Yu-Jiang Bi$^{2,3}$, Terrence Draper$^{4}$, Keh-Fei Liu$^{4}$, Zhaofeng Liu$^{2,5}$, Yi-Bo Yang$^{1,6,7,8}$
\vspace*{-0.5cm}
\begin{center}
\large{
\vspace*{0.4cm}
\includegraphics[scale=0.15]{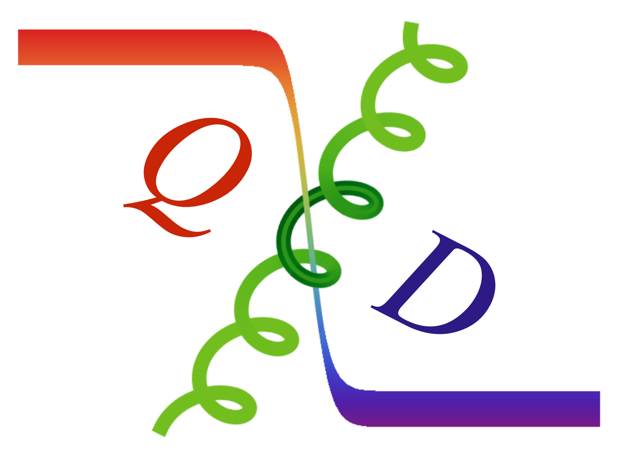}\\
\vspace*{0.4cm}
($\chi$QCD Collaboration)
}
\end{center}
}
\affiliation{
$^{1}$\mbox{CAS Key Laboratory of Theoretical Physics, Institute of Theoretical Physics, Chinese Academy of Sciences, Beijing 100190, China}\\
$^{2}$\mbox{Institute of High Energy Physics, Chinese Academy of Sciences, Beijing 100049, China}\\
$^{3}$\mbox{Tianfu Cosmic Ray Research Center, Chengdu 610213, China} \\
$^{4}$\mbox{Department of Physics and Astronomy, University of Kentucky, Lexington, KY 40506, USA}\\
$^{5}$\mbox{Center for High Energy Physics, Peking University, Beijing 100871, China}\\
$^{6}$\mbox{School of Fundamental Physics and Mathematical Sciences, Hangzhou Institute for Advanced Study, UCAS, Hangzhou 310024, China}\\
$^{7}$\mbox{International Centre for Theoretical Physics Asia-Pacific, Beijing/Hangzhou, China}\\
$^{8}$\mbox{School of Physical Sciences, University of Chinese Academy of Sciences,
Beijing 100049, China}\\
}

\begin{abstract}
We present the vector, scalar and tensor renormalization constants (RCs) using overlap fermions with either regularization independent momentum subtraction (RI/MOM) or symmetric momentum subtraction (RI/SMOM) as the intermediate scheme on the lattice with lattice spacings $a$ from 0.04 fm to 0.12 fm. Our gauge field configurations from the MILC and RBC/UKQCD collaborations include sea quarks using either the domain wall or the HISQ action, respectively. The results show that RI/MOM and RI/SMOM can provide consistent renormalization constants to the $\overline{\textrm{MS}}$
 scheme, after proper $a^2p^2$ extrapolations. But at $p\sim 2$\,GeV, both RI/MOM and RI/SMOM suffer from nonperturbative effects which cannot be removed by the perturbative matching.  The comparison between the results with different sea actions also suggests that the renormalization constant is discernibly sensitive to the lattice spacing but not to the bare gauge coupling in the gauge action.
\end{abstract}

\maketitle

\section{Introduction}\ 

Lattice QCD has been shown to be a very powerful and accurate framework for predicting the hadron spectrum, since the spectrum is scale and scheme independent, and can provide reliable and model-independent predictions once systematic uncertainties such as finite volume, lattice spacing, quark masses, and QED effects are under control. For scale dependent quantities, such as physical quark masses, chiral condensates, parton distribution functions and so on, the situation is more nuanced as these quantities are typically considered under the $\overline{\textrm{MS}}$ renormalization procedure in phenomenological studies, and then are generally different from the bare quantities one can determine nonperturbatively from lattice QCD.
When the lattice spacing $a$ is small enough and so the ultraviolet cutoff $1/a$ is large, the difference is due to that of lattice regularization and dimensional regularization, which can be calculated and matched perturbatively. Such a matching is equivalent to calculating the same vertex correction in a regularization independent (RI) scheme under both regularizations, and taking the ratio. In such a sense, the calculation under lattice regularization can be done with a nonperturbative simulation, followed by a perturbative calculation under dimensional regularization.

These are various choices for such a regularization independent scheme. A straightforward choice is the regularization independent momentum subtraction (RI/MOM) scheme, which considers the vertex correction in the forward off-shell parton state~\cite{Martinelli:1994ty}. The corresponding perturbative calculation is relatively simple and those for the quark bilinear currents with all 16 gamma matrices have been obtained at 3 loops or more~\cite{Franco:1998bm, Chetyrkin:1999pq,Gracey:2003yr,Gracey:2022vqr,Gracey:2022vjc}. But the singularity from the zero momentum transfer at the current position can create nonperturbative effects and poor perturbative convergence, especially for the scalar current renormalization constant $Z_S$ which is crucial for the quark mass and chiral condensate determination. Eventually it can cause more than 1\% uncertainty on $Z_S$ which is difficult to suppress despite a better nonperturbative lattice QCD calculation.

In view of this, Refs.~\cite{Aoki:2007xm,Sturm:2009kb} proposed the symmetric momentum subtraction (RI/SMOM) scheme, which requires that the momentum transfer at the current is the same as that on the external parton leg. With such a setup, the perturbative convergence has been verified to be better than that of $\MOM$ scheme up to the 3-loop level for the scalar current, and the nonperturbative effects such as the $1/m_q^n$ poles in the scalar and pseudoscalar currents disappear~\cite{Sturm:2009kb}. Thus the RI/SMOM scheme seems to be helpful in suppressing the systematic uncertainty of $Z_S$ to the $\sim$ 0.2\% level, and has been widely used in quark mass determinations, i.e., \cite{RBC:2010qam,RBC:2014ntl,Lytle:2018evc}.

In our previous study~\cite{Bi:2017ybi}, we calculated $Z_S$ using overlap fermions on 2+1 domain wall fermion (DWF) gauge ensembles at $a=0.114$ fm with physical pion mass, through both the RI/MOM and RI/SMOM schemes. With 1-step of HYP smearing applied on the fermion action, we found that it is impossible to find a good $a^2p^2$ extrapolation window to remove the discretization error of $Z_S^{\overline{\textrm{MS}}}$(2\,GeV) in the RI/SMOM case. There is also a recent work which found a 10--20\% discrepancy between the $Z_S$ through the RI/MOM and RI/SMOM schemes using the clover fermion at $a$=0.116 and 0.093 fm, even though the discrepancy decreases with the lattice spacing~\cite{Hasan:2019noy}. Thus, a more careful comparison of the RI/MOM and RI/SMOM schemes with a larger range of lattice spacings is warranted for accurate hadron structure studies in the future. 

The setup of our calculation, including the detailed definition of RI/MOM and RI/SMOM, and fermion and gauge actions, will be presented in Sec.~\ref{sec:setup}. Sec.~\ref{sec:norm} provides a study on the quark self-energy definition and the vector/axial-vector current normalization, and the details of the systematic uncertainty analysis are given in Sec.~\ref{sec:case}. The results at lattice spacings $a$ from 0.04 fm to 0.20 fm using either the DWF sea or the highly improved staggered quark (HISQ) sea are presented in Sec.~\ref{sec:summary}.

\section{Numerical setup}\label{sec:setup}

In this work, we use overlap fermions~\cite{{Narayanan:1994gw,Neuberger:1997fp,Liu:2002qu}} as valence quarks to calculate the renormalization constants (RCs) in the $\MOM$ and $\SMOM$ schemes. Overlap fermions have perfect chiral symmetry which guarantees that $Z_{\text{P}}=Z_\text{S}$ and $Z_\text{A}=Z_\text{V}$ when the nonperturbative effects in the IR region are {removed properly}.  
The overlap Dirac operator is written as 
\begin{eqnarray}
D_{ov}(\rho)=\rho(1+\gamma_5\epsilon(\gamma_5D_w(-\rho))),
\end{eqnarray}
where $D_w(-\rho)$ is the Wilson fermion operator. $\rho$ is the mass parameter and is chosen to be $\rho=1.5$ in our calculation. $\epsilon$ is the sign function and satisfies $\epsilon^2=1$. One can easily find that $D_w(\rho)$ satisfies the Ginsparg-Wilson relation~\cite{Ginsparg:1981bj},
\begin{eqnarray}
D_{ov}\gamma_5+\gamma_5D_{ov}=\frac{1}{\rho}D_{ov}\gamma_5D_{ov},
\end{eqnarray}
and the effective Dirac operator is defined as
\begin{eqnarray}
D_c=\frac{ D_{ov}}{1-\frac{1}{2\rho}D_{ov}},
\end{eqnarray}
{which} satisfies $\{D_c,\gamma_5\}=0$~\cite{Chiu:1998gp}.
The massive effective inverse Dirac operator is
$D_c(m)^{-1} = 1/(D_c-m)$ which has the same form as that in the continuum~\cite{Liu:2002qu}.

\begin{table}[htbp]
  \centering
  \begin{tabular}{l|lcrlc}
  \toprule
tag &  $6/g^2$ & $L$ & $T$ & $a(\mathrm{fm})$ & $m_{\pi}$ (MeV) \\
\hline
24D &  1.633 & 24 & 64 & 0.194(2) & 139 \\
\hline
24DH & 1.633 & 24 & 64 & 0.194(2) & 337  \\
\hline
32Dfine & 1.75  & 32 & 64 & 0.143(2) & 139  \\
\hline
48I &  2.13 & 48 & 96 & 0.1141(2) &139  \\
\hline
24I &  2.13 & 24 & 64 & 0.1105(2) & 340/432/576/693  \\
\hline
64I &  2.25 & 64 & 128 & 0.0837(2) &139\\
\hline
48If &  2.31 & 48 & 96 & 0.0711(3) &280 \\
\hline
32If &  2.37 & 32 & 64 & 0.0626(4) & 371 \\
\hline
HISQ12L &  3.60 & 48 & 64 & 0.1213(9) & 130\\
\hline
HISQ12H &  3.60 & 24 & 64 & 0.1213(9) & 310\\
\hline
HISQ09L &  3.60 & 64 & 96 & 0.0882(7) & 130\\
\hline
HISQ09H &  3.78 & 32 & 96 & 0.0882(7)& 310\\
\hline
HISQ06 &  4.03 & 48 & 144 & 0.0574(5)& 310 \\
\hline
HISQ04 &  4.20 & 64 & 192 & 0.0425(4)& 310 \\
\hline
  \end{tabular}
  \caption{Setup of the ensembles, including the bare coupling constant $g$, lattice size $L^3\times T$, lattice spacing $a$ and sea pion mass $m_{\pi}$. }
  \label{tab:lattice}
\end{table}

In our calculations, we use two sets of dynamical gauge configurations, namely those with 2+1 flavor domain wall fermions (DWF)~\cite{Kaplan:1992bt} with the Iwasaki gauge action from the RBC/UKQCD collaboration~\cite{RBC:2012cbl,Blum:2014tka,Mawhinney:2019cuc} and those with 2+1+1 flavor highly improved staggered quarks (HISQ)~\cite{Kogut:1974ag,Follana:2006rc} with the Symanzik gauge action from the MILC collaboration~\cite{MILC:2010pul,Bazavov:2012xda,Bazavov:2017lyh}. The information of the ensembles we use in this work can be found in Table~\ref{tab:lattice}. 

The RCs in different renormalization schemes can be obtained through imposing the specific renormalization conditions on the bare amputated Green's functions. If one uses a point source in a lattice simulation, then the bare Green's function $G_\mathcal{O}$ can be defined as
\begin{eqnarray}
G_\mathcal{O}(p_1,p_2)=\sum_{x,y}e^{-i(p_1\cdot x-p_2\cdot y)}\langle
\psi(x)\mathcal{O}(0)\bar{\psi}(y)\rangle,
\end{eqnarray}
where $\mathcal{O}=\bar{\psi}\Gamma\psi$ is the quark operator and the interpolation gamma matrix $\Gamma$ is chosen to be $I$, $\gamma_5$, $\gamma_\mu$,  $\gamma_\mu\gamma_5$ and $\sigma_{\mu\nu}$ for the scalar (S), pseudoscalar (P), vector (V),
axial-vector(A) or tensor (T) currents, respectively. Then the amputated Green function can be obtained by
\begin{eqnarray}
\Lambda_\mathcal{O}(p_1,p_2)=S^{-1}(p_1)G_\mathcal{O}(p_1,p_2)S^{-1}(p_2),
\end{eqnarray}
where $S(p)$ is the {point source} quark propagator 
\begin{eqnarray}
S(p)=\sum_xe^{-ip\cdot x}\langle \psi(x)\bar{\psi}(0)\rangle.
\end{eqnarray}

According to the LSZ reduction, one can define the renormalized amputated Green's function as 
\begin{eqnarray}\label{eq:lsz}
\Lambda^r_{\mathcal{O},R}(p_1,p_2)=\frac{Z^r_\mathcal{O}}{Z^r_q}\Lambda_{\mathcal{O},B}(p_1,p_2),
\end{eqnarray}
where $r\in\{\MSbar$,~\MOM,~\SMOM$\}$ represents one of the different renormalization schemes we will consider in this work, the $Z^r$ are the RCs of the quark field and operators, and subscripts R and B represent the renormalized and bare quantities, respectively,
\begin{eqnarray}
S^r_R=Z_q^rS^r_B,\
\mathcal{O}^r_R=Z_\mathcal{O}^r\mathcal{O}_B.
\end{eqnarray}

In the $\MOM$ scheme, the RCs can be determined by the following renormalization conditions~\cite{Martinelli:1994ty},
\begin{subequations}
\label{eq:rimom}
\begin{eqnarray}
\label{eq:rimom_prop}
Z_q^{\MOM}&=&\mathop{\textrm{lim}}\limits_{m_R\rightarrow0}\frac{-i}{48}\textrm{Tr}\Big[\gamma_\mu\frac{\partial S_B^{-1}(p)}{\partial p_\mu}\Big]_{p^2=\mu^2},  \\
\label{eq:rimom_Zo}
Z^\MOM_{\mathcal{O}}&=&
\mathop{\text{lim}}\limits_{m_R\rightarrow0}\frac{Z_q^\MOM}{\frac{1}{12}\text{Tr}[\Lambda_{\mathcal{O},B}(p,p)\Lambda^{\text{tree}}_{\mathcal{O}}(p,p)^{-1}]_{p^2=\mu^2}},\non\\
\end{eqnarray}
\end{subequations}
where the momenta of the external quark legs are 
chosen to satisfy $p_1=p_2=p$ and $\mu^2=p^2$ is the renormalization scale of {the} $\MOM$ scheme.
However, the definition in Eq.~(\ref{eq:rimom_prop}) needs to calculate the derivative with respect to the momentum, and this will inevitably introduce a systematic error since the momentum is discrete in lattice simulations. A more convenient method to obtain $Z_q^\MOM$ is using the vector vertex correction
\begin{equation}\label{eq:rimom_vex}
Z_q^{\MOM}=\mathop{\text{lim}}\limits_{m_R\rightarrow0}\frac{Z_V}{48}\text{Tr}[\Lambda^\mu_{V,B}(p)\gamma_\mu]_{p^2=\mu^2}.
\end{equation}
It is easy to verify that Eq.~(\ref{eq:rimom_prop}) and Eq.~(\ref{eq:rimom_vex}) are equivalent by using the Ward identity~\cite{Martinelli:1994ty},
\begin{equation}\label{eq:WI1}
Z_V\Lambda_{V,B}(p)=-i\frac{\partial S^{-1}_B(p)}{\partial p_\mu}.
\end{equation}
A modified version of the RI/MOM scheme is the $\RIpMOM$ scheme, which replaces $Z_q^{\MOM}$ by $Z_q^{\RIp}$~\cite{Gracey:2003yr},
\begin{align}\label{eq:riprime_Zq}
  Z_q^{\RIp}= \mathop{\textrm{lim}}\limits_{m\rightarrow0}\frac{-i}{12p^2}\textrm{Tr}\Big[S^{-1}_B(p)\slashed{p}\Big]_{p^2=\mu^2}.
\end{align}
Based on the Lorentz structure of the vector current vertex correction, one can obtain another expression of $Z_q^{\RIp}$ through the transverse projection on the forward vector current,
\begin{equation}\label{eq:Zq_RIp_vex}
Z_q^{\RIp, {\rm ver}}=\mathop{\text{lim}}\limits_{m\rightarrow0}\frac{Z_V}{36}\text{Tr}[\Lambda^\mu_{V,B}(p)(\gamma_\mu-\frac{\slashed{p}p_\mu}{p^2})].    
\end{equation}

On the other hand, RI/SMOM is an alternative nonperturbative renormalization scheme~\cite{Aoki:2007xm,Sturm:2009kb}. In this scheme, the momenta of external quark legs are symmetrically set to be 
\begin{eqnarray}
p_1^2=p_2^2=(p_2-p_1)^2=\mu^2{.}
\end{eqnarray}
The renormalization conditions for scalar, pseudoscalar and tensor currents are similar to those in the RI/MOM scheme. The renormalization conditions for the quark self energy and quark bilinear operators are chosen to be 
\begin{subequations}
\label{eq:rismom}
\begin{eqnarray}
\label{eq:rismom_Zq}
Z_q^{\SMOM}&=&Z_q^{\RIp},  \\
\label{eq:rismom_ZS}
Z_{\mathcal{O}=S/P/T}^\SMOM&=&\mathop{\text{lim}}\limits_{m\rightarrow0}\frac{Z_q^\SMOM}{\frac{1}{12}\text{Tr}[\Lambda_{\mathcal{O},B}(p_1,p_2)\Lambda^{\text{tree}}_{\mathcal{O}}(p_1,p_2)^{-1}]},\non\\
\label{eq:rismom_ZV}
Z_V^\SMOM&=&\mathop{\text{lim}}\limits_{m\rightarrow0}\frac{Z_q^\SMOM}{\frac{1}{12q^2}\text{Tr}[q_\mu\Lambda^\mu_{V,B}(p_1,p_2)\slashed{q}]},\\
\label{eq:rismom_ZA}
Z_A^\SMOM&=&\mathop{\text{lim}}\limits_{m\rightarrow0}\frac{Z_q^\SMOM}{\frac{1}{12q^2}\text{Tr}[q_\mu\Lambda^\mu_{A,B}(p_1,p_2)\gamma_5\slashed{q}]},
\end{eqnarray}
\end{subequations}
Eq.~(\ref{eq:rismom_Zq}) is the definition of the RC of the quark field strength in
the $\RIp$ scheme~\cite{Gracey:2003yr}. 
Using Eq.~(\ref{eq:rismom_ZV}), one can obtain the $Z_q$ in the $\SMOM$ scheme by 
\begin{equation}\label{eq:Zq_smom_vex}
Z_q^{\SMOM, {\rm ver}}=\mathop{\text{lim}}\limits_{m_R\rightarrow0}\frac{Z_V}{12q^2}\text{Tr}[q_\mu\Lambda^\mu_{V,B}(p_1,p_2)\slashed{q}].
\end{equation}
In the perturbative theory, the bare quark propagator can be written in the generic form
\begin{equation}\label{eq:bare_prop}
-iS_B^{-1}(p)=\slashed{p}\Sigma_1(p^2)-m_0\Sigma_2(p^2)
\end{equation}
Then one can verify that $Z_q^{\SMOM, \rm ver}$ from Eq.~(\ref{eq:Zq_smom_vex}) and $Z_q^{\RIp}$ defined in Eq.~(\ref{eq:riprime_Zq}) are equivalent in the continuum limit
using the Ward
identity~\cite{Karsten:1980wd,Bochicchio:1985xa},
\begin{equation}
Z_Vq_\mu\Lambda^\mu_{V,B}(p_1,p_2)=-i\Big(S_B^{-1}(p_2)-S_B^{-1}(p_1)\Big),    
\end{equation}
where we have used the relation $\Sigma_{1,2}(p_1^2)=\Sigma_{1,2}(p_2^2)=\Sigma_{1,2}(q^2)$ since the momentum is chosen to be symmetric in the $\SMOM$ scheme.
In Sec.~(\ref{sec:zq}), we will show that the discretization error would be larger if one chose the definition Eq.~(\ref{eq:riprime_Zq}) to calculate $Z_q^{\SMOM}$. So {a} better choice is  using the vector vertex correction {in Eq.~(\ref{eq:Zq_smom_vex})} to calculate the RC of the quark field strength. 

An RC in the intermediate scheme $r\in\{\MOM,\SMOM\}$ can be converted to the $\MSbar$ scheme through multiplication by the matching factor $C_{q/{\mathcal{O}}}^{\MSbar,r}$,
\begin{eqnarray}
Z_{q/{\mathcal{O}}}^{\MSbar}(\mu)=C_{q/{\mathcal{O}}}^{\MSbar,r}(\mu)Z_{q/\mathcal{O}}^{r}(\mu).
\end{eqnarray}
More precisely, the matching factor $C_{q}^\MOM$ and the ratio of $C_{\mathcal{O}}^\MOM$ to $C_{q}^\MOM$ can be obtained through the following way, 
\begin{eqnarray}\label{eq:match_factor}
C_q^{\MSbar,\MOM}(\mu)&=&\mathop{\textrm{lim}}\limits_{m_R\rightarrow0}\frac{48i}{\text{Tr}\Big[\gamma_\mu\frac{\partial (S^{\MSbar}_R)^{-1}(p)}{\partial p_\mu}\Big]_{p^2=\mu^2}},  \\
C_{\mathcal{O}}^{\MSbar,\MOM}(\mu)&=&\mathop{\text{lim}}\limits_{m_R\rightarrow0}\frac{C_q^{\MSbar,\MOM}(\mu)}{12}\nonumber\\&&\text{Tr}[\Lambda^{\MSbar}_{\mathcal{O},R}(p,p)\Lambda^{\text{tree}}_{\mathcal{O}}(p,p)^{-1}]_{p^2=\mu^2}.\nonumber\\
\end{eqnarray}
The above equations can be easily obtained through replacing $S_B(p)$ and $\Lambda_{\mathcal{O},B}(p)$ in Eq.~(\ref{eq:rimom}) by $S_R^\MSbar(p)$ and $\Lambda^{\MSbar}_{\mathcal{O},R}(p,p)$.
The matching factors between the $\SMOM$ and $\MSbar$ schemes can be determined in a similar way. 

When using the $\MOM$ and $\SMOM$ schemes to calculate RCs at a specific lattice spacing $a$, the window of the renormalization scale $\mu$ is chosen to be 
\begin{eqnarray}
\Lambda_{QCD}\ll \mu \ll C\pi/a,
\end{eqnarray}
where the nonperturbative effects from chiral symmetry breaking and ultraviolet effects caused by the lattice spacing are highly suppressed in such a window, and $C$ is an unknown constant which is sensitive to the regularization and renormalization schemes.  

In our calculation, the boundary conditions of all four directions are chosen to be periodic. {For a discretized momentum on the lattice,
\begin{equation}
ap=2\pi\Big(\frac{k_1}{L},\frac{k_2}{L},\frac{k_3}{L},\frac{k_4}{T}\Big),    
\end{equation}
the leading order discretization error will be proportional to the  ``democratic"  factor
\begin{eqnarray}\label{eq:demo_cond}
c(p)\equiv \frac{p^{[4]}}{(p^2)^2},  \text{where}~ p^{[4]}=\sum\limits_\mu p^4_\mu,~p^2=\sum\limits_{\mu}p^2_\mu.
\end{eqnarray}}

In the MOM scheme, we choose $c(p)<0.28$ for all the momenta used in the calculation, but the momentum constraint Eq.~(\ref{eq:demo_cond}) in the SMOM scheme can only allow us to choose the momenta with much larger $c(p)$, such as the ones used in this work, namely $ap_1$=($q$,$q$,0,0) and $ap_2$=(0,$q$,$q$,0) for which $c(p)=0.5$.

\section{Quark self energy and normalization}\label{sec:norm}

\subsection{Definition of $Z_q$}\label{sec:zq}

\begin{figure}[]
	\includegraphics[scale=0.6]{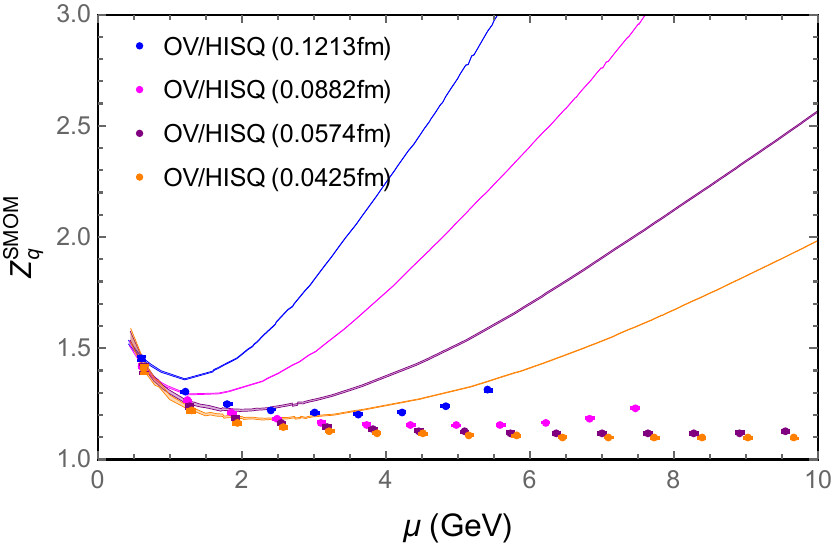}
	\includegraphics[scale=0.6]{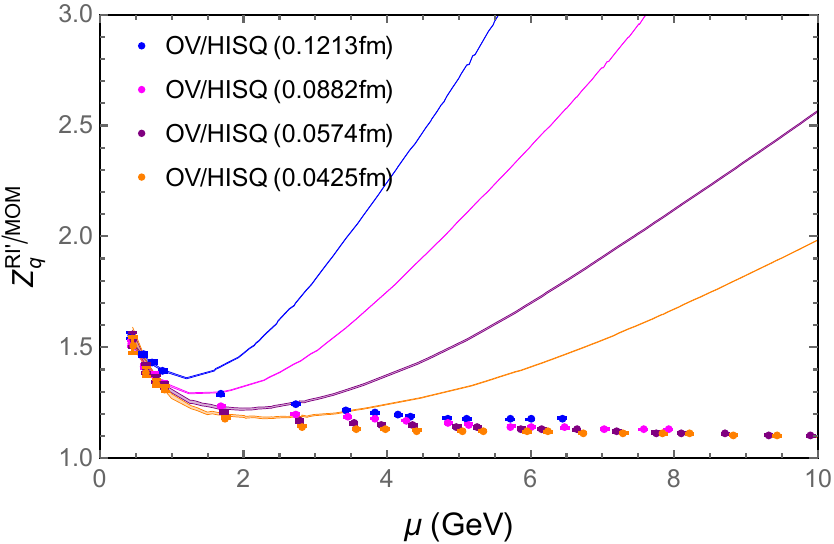}
\caption{The comparison of $Z_q$ using the overlap fermion on the HISQ ensemble (OV/HISQ for short) obtained by the bare quark propagator and vector current correction. 
The data points in the upper panel and lower panel are obtained by Eq.~(\ref{eq:Zq_smom_vex}) ($Z_q^{\SMOM, {\rm ver}}= \mathop{\text{lim}}\limits_{m_R\rightarrow0}\frac{Z_V}{12q^2}\text{Tr}[q_\mu\Lambda^\mu_{V,B}(p_1,p_2)\slashed{q}]$), and Eq.~(\ref{eq:Zq_RIp_vex}) ($Z_q^{\RIp, {\rm ver}}=\mathop{\text{lim}}\limits_{m\rightarrow0}\frac{Z_V}{36}\text{Tr}[\Lambda^\mu_{V,B}(p)(\gamma_\mu-\frac{\slashed{p}p_\mu}{p^2})]$), respectively. The curves with the four colors correspond to the $Z_q^{\RIp}
\equiv \mathop{\textrm{lim}}\limits_{m\rightarrow0}\frac{-i}{12p^2}\textrm{Tr}\Big[S^{-1}_B(p)\slashed{p}\Big]_{p^2=\mu^2}$ defined in Eq.~(\ref{eq:riprime_Zq}). The curve and data points with the same color correspond to results at the same lattice spacing, which is specified within parentheses in the legends.
The data points in the lower panel show better convergence in the continuum extrapolation, since the momentum in the $\RIp$ scheme (lower panel) is chosen to be body diagonal $\sim 1/2(p,p,p,p)$ in the forward case while in the $\SMOM$ scheme (upper panel) it is chosen to be $\sqrt{2}/2(0,0,p,p)$. Thus the discretization error, as indicated by the convergence of results for the different ensembles, in the lower panel is smaller that in the upper panel.
}
\label{fig:Zq_compa}
\end{figure}

We present the results of $Z_q^{\SMOM, {\rm ver}}$ (data points) from Eq.~(\ref{eq:Zq_smom_vex}) and $Z_q^{\SMOM}=Z_q^{\RIp}$ (curves) from  Eq.~(\ref{eq:riprime_Zq}) in the upper panel of Fig.~\ref{fig:Zq_compa}. The lower panel  of Fig.~\ref{fig:Zq_compa} shows the comparison of $Z_q^{\RIp}$ (curves)  to the alternative vertex correction version in the MOM case, $Z_q^{\RIp, {\rm ver}}$ (data points) from Eq.~(\ref{eq:Zq_RIp_vex}). The blue, {magenta, purple, and orange dots} represent the results on the HISQ12, HISQ09, HISQ06 and HISQ04 ensembles, respectively.  One can see that the deviations between the quark propagator and vertex definitions increase with $\mu$ and decrease with $a$; thus such a behavior agrees with our expectation of a discretization error. Both $Z_q^{\RIp, {\rm ver}}$ and $Z_q^{\SMOM, {\rm ver}}$ have better convergence in the continuum extrapolation, compared to the original $ Z_q^{\RIp}$ defined from the quark propagator. The discretization error of $Z_q^{\RIp, {\rm ver}}$ is even smaller than that of $Z_q^{\SMOM, {\rm ver}}$, since the momentum used in the MOM case is closer to the body diagonal direction and thus has a smaller $c(p)$. 

\begin{figure}[htbp]
	\centering
	\includegraphics[width=0.48\textwidth]{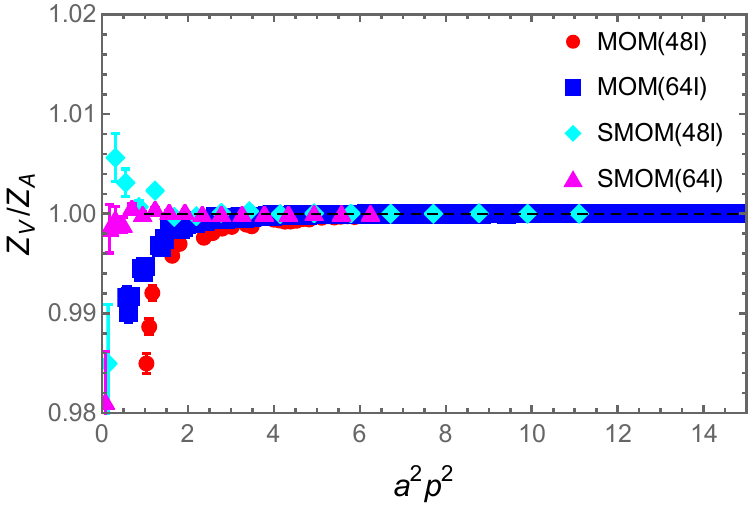}
	\caption{$Z_V/Z_A$ in the chiral limit under the $\MOM$ and $\SMOM$ schemes on the 48I and 64I ensembles.}
	\label{fig:ZVZA}
\end{figure}

In the $\MOM$ and $\SMOM$ schemes, the ratio $Z_V/Z_A$ can be obtained by the following ratios,
\begin{eqnarray}
\frac{Z_V^\MOM}{Z_A^\MOM}&=&\frac{\text{Tr}[\Lambda^\mu_{A,B}(p)\gamma_5\gamma_\mu]}{\text{Tr}[\Lambda^\mu_{V,B}(p)\gamma_\mu]}\Big|_{p^2=\mu^2},  \non\\
\frac{Z_V^\SMOM}{Z_A^\SMOM}&=&\frac{\text{Tr}[q_\mu\Lambda^\mu_{A,B}(p_1,p_2)\gamma_5\slashed{q}]}{\text{Tr}[q_\mu\Lambda^\mu_{V,B}(p_1,p_2)\slashed{q}]}\Big|_{p_1^2=p_2^2=q^2},
\end{eqnarray}
In Fig.~\ref{fig:ZVZA}, we plot the ratios $Z_V/Z_A$ from the $\MOM$ and $\SMOM$ schemes, which have been linearly extrapolated to the chiral limit. The ratio in $\MOM$ is consistent with 1 in the large $a^2p^2$ region but deviates from 1 when $a^2p^2$ is small due to the effect of the Goldstone mass pole in the forward axial vector current~\cite{Martinelli:1994ty}. Comparing the results on the 48I and 64I ensembles (red filled circles and blue filled boxes), one can see this 
nonperturbative effect occurs at a smaller $a^2p^2$ region when the lattice spacing is smaller. On the other hand, the effects on the 48I (cyan diamonds) and 64I (magenta triangles) are highly suppressed in the $\SMOM$ case, as the value of $Z_V/Z_A$ is consistent with 1 at much lower momentum scale.

\subsection{Normalization of axial vector current}

\begin{figure}[htbp]
	\centering
	\includegraphics[width=0.48\textwidth]{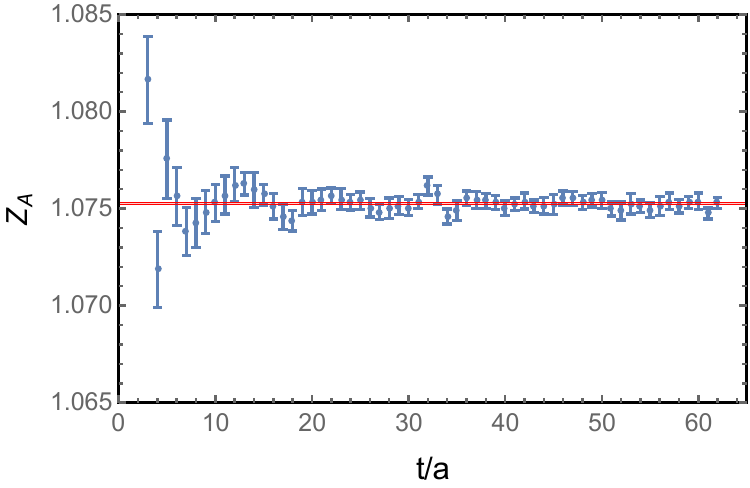}
	\caption{The result of $Z_A$ on the 64I ensemble obtained with the PCAC relation versus source and sink time separation $t/a$ using Eq.~(\ref{eq:PCAC}). The quark mass is $m_qa=0.012$.}
	\label{fig:ZA64I}
\end{figure}

The normalization constant of the axial vector current can be calculated through the PCAC relation,
\begin{equation}
Z_A\partial_\mu A_\mu=2Z_mZ_pm_qP,    
\end{equation}
with $Z_mZ_p$=1 for overlap fermions. The PCAC relation can be changed into~\cite{Bi:2017ybi}
\begin{eqnarray}
Z_A=\frac{2m_q\sum_{\vec{x}}\langle\Omega|P(t,\vec{x})P^\dag(0)|\Omega\rangle}{\sum_{\vec{x}}\langle\Omega|\partial_\mu A_\mu(t,\vec{x})P^\dag(0)|\Omega\rangle}|_{t\rightarrow \infty}.
\end{eqnarray}
Then we have 
\begin{eqnarray}\label{eq:PCAC}
Z_A=\frac{4am_qC_{PP}(t)}{C_{A_4P}(t-a)-C_{A_4P}(t+a)}|_{a\rightarrow 0},
\end{eqnarray}
where $C_{PP}(t)$ and $C_{A_4P}(t)$ are the two-point correlation functions of pseudoscalar-pseudoscalar operators and pseudoscalar-axial vector operators, respectively. In Fig.~\ref{fig:ZA64I}, we show the {64I} result of $Z_A$ obtained through the PCAC relation; the valence quark mass equals $m_va=0.012$. One can see the statistical uncertainty of $Z_A$ is quite small and the plateau is stable in the region $t\in[6a,60a]$. We obtain $Z_A$=1.0753(1) using a constant fit.

\begin{table*}[htbp]
  \centering
  \begin{tabular}{c|cccccccccccc}
  \toprule
Ensemble & 24D & 24DH & 32Dfine &   48I & 64I & 48If & 32If  \\
\hline
$Z_A$ & 1.2186(4)(1) & 1.2240(3)(1) &  1.1417(2)(1) & 1.1036(1)(1) & 1.0787(1)(1) &  1.0698(1)(1) & 1.0649(2)(2)  \\
\hline
Ensemble & HISQ12L & HISQ12H & HISQ09L & HISQ09H & HISQ06 & HISQ04 \\
\hline
$Z_A$ &  1.1088(3)(1) & 1.1092(2)(1) & 1.0822(1)(1) & 1.0832(1)(1) & 1.0617(1)(1)  & 1.0519(1)(1) \\
\hline
  \end{tabular}
  \caption{The axial current renormalized constants of different ensembles. The values in the two brackets following the central value correspond to the statistical error and the systematic error caused by the finite volume effect.}
  \label{tab:ZA}
\end{table*}

\begin{figure}[htbp]
	\centering
	\includegraphics[width=0.48\textwidth]{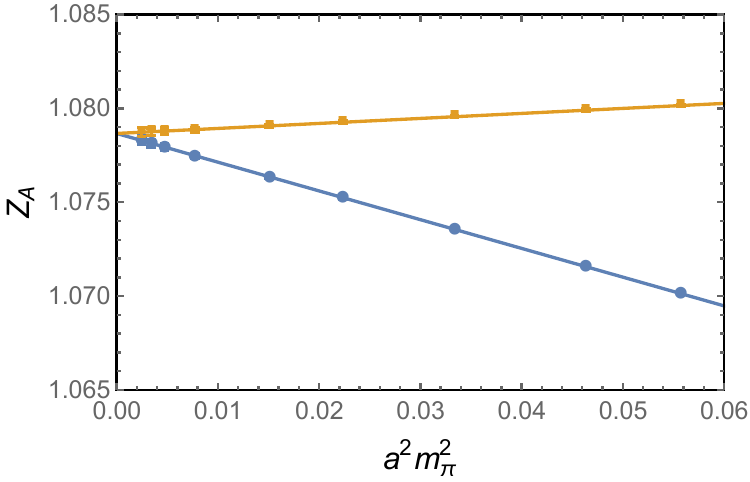}
	\caption{The renormalization constant $Z_A$ versus the square of the pion mass on the 64I ensemble. The orange and blue data represent the results with and without the subtraction of the $-a^2m_{\pi}^2/6$ term in Eq.~(\ref{eq:ZA_chiralextra}) from the data, respectively.}
	\label{fig:ZA64I1}
\end{figure}

In Eq.~(\ref{eq:PCAC}), we replace the derivative by the difference, so there is an additional discretization error. To make it clear, we consider the expression of $C_{PP}(t)$ and $C_{A_4 P}(t)$ at a large time separation,
\begin{eqnarray}\label{eq:exp2pt}
C_{PP}(t)&=&Ae^{-m_\pi t}+Ae^{-m_\pi (T-t)},\nonumber\\
C_{A_4P}(t)&=&Be^{-m_\pi t}-Be^{-m_\pi (T-t)}.
\end{eqnarray}
Substituting Eq.~(\ref{eq:exp2pt}) into Eq.~(\ref{eq:PCAC}), one can obtain
\begin{align}\label{eq:ZA_chiralextra}
Z_A(m_q)&=\frac{4aAm_q}{B[e^{am_\pi}-e^{-am_\pi}]}\nonumber\\
&=\frac{2Am_q}{Bm_\pi}\left(1-\frac{a^2m^2_\pi}{6}+O(a^4m_\pi^4)\right).
\end{align}
In Fig.~\ref{fig:ZA64I1}, we plot $Z_A$ versus valence quark mass, with (orange points) and without (blue points) the subtraction
of the $-a^2m^2_\pi/6$ term from the data. It is clear that $Z_A$ changes significantly with the quark mass $m_q$ due to the $a^2m^2_\pi/6$ term, while the $m_q$ dependence is much milder when that term is subtracted. 

Then we use a linear $m_{\pi}^2$ extrapolation to obtain $Z_A=1.07867(6)$ in the chiral limit. The final results of $Z_A$ on different ensembles are shown in Table~\ref{tab:ZA}. In order to estimate the finite volume effect, we dropped the lightest two quark masses, which have a relatively larger finite volume effect. Then we obtain that $Z_A=1.07865(6)$; this value is very close to the result extrapolated with all quark masses, with the systematic error caused by the finite volume less than $0.01\%$. We list this systematic error in Table~\ref{tab:ZA}.

Another issue we would like to discuss here is the quark mass dependence of $Z_A$.
After subtracting the $a^2m^2_\pi/6$ term, the $m_q$ dependence is still nonvanishing as shown in the orange data of Fig.~\ref{fig:ZA64I1}, with $\frac{\partial Z_A}{\partial m_q}=0.020(1)~\mathrm{GeV}^{-1}$ on the 64I ensemble. The slopes on all ensembles we studied in this work are consistent with an estimate $0.10(3)\,\mathrm{GeV}\times a^2 \,$ within the uncertainty, which thus looks like a discretization error.

{We also calculate $Z_A$ on the four 24I ensembles at the same lattice spacing $a=0.11$~fm but different light sea quark masses from $m_la=$~0.005 to 0.03, and list the results in Table~\ref{tab:sea_dep_ZA}. With larger statistics compared to the previous studies~\cite{Liu:2013yxz,Wang:2016lsv}, we can extract a nonzero sea quark mass dependence with
the following ansatz
\begin{equation}\label{eq:extra_sea}
Z(m_l^{{\rm sea}, R})=Z(0)+c m_l^{{\rm sea}, R},
\end{equation}
where $m_l^{{\rm sea}, R}\equiv Z^{sea}_m(m^{sea}_l+m_{res})$ is the renormalized light sea quark mass of the domain wall fermion used in the RBC/UKQCD gauge ensemble, $m_{res}$ is the residual quark mass of the domain wall fermion, and $Z^{sea}_m=1.578(2)$~\cite{RBC:2010qam} is the renormalization constant of the sea quark mass. The slope we get is $c=0.040(4)\,\GeV^{-1}$ and is consistent with the valence quark mass dependence on the 24I ensemble with the lightest quark mass, which equals $0.040(2)\,\GeV^{-1}$. We speculate the $m_l^{{\rm sea}}$ dependence on the ensembles at the other lattice spacings is also a discretization error at the same order of the valence quark mass $m_q$ dependence.
}

\begin{table}[htbp]
  \centering
  \begin{tabular}{c|cccc}
  \toprule
$m_la$ & 0.005 & 0.01 & 0.02 & 0.03  \\
\hline
$Z_A$ & 1.1020(2) & 1.1023(2) & 1.1036(2) & 1.1047(2) \\
\hline
\hline
  \end{tabular}
\caption{$Z_A$ of four different 24I ensembles. The first row lists the light sea quark masses of these ensembles. Using Eq.~(\ref{eq:extra_sea}) to do the linear extrapolation of light sea quark mass, we obtain that the slope $c$ for $Z_A$ is 0.040(4)\,GeV$^{-1}$.}
  \label{tab:sea_dep_ZA}
\end{table}

\section{Case studies}\label{sec:case}

{As shown in the previous section, the $Z_q$ defined from the vector current vertex correction and quark propagator are consistent with each other for both the SMOM and MOM cases, while $Z_A/Z_V$ can differ from unity obviously at small $a^2p^2$ in the MOM case. Thus in this section, we will consider the ratio $Z_X/Z_V$ instead of $Z_X/Z_A$ to avoid the nonperturbative effect in the axial-vector current vertex corrections in the MOM case. Most of the discussions in this section are based on the physical pion mass ensemble 64I; the procedure is similar on the other ensembles.}

\subsection{Quark field renormalization}

\begin{figure}[]
	\includegraphics[scale=0.7]{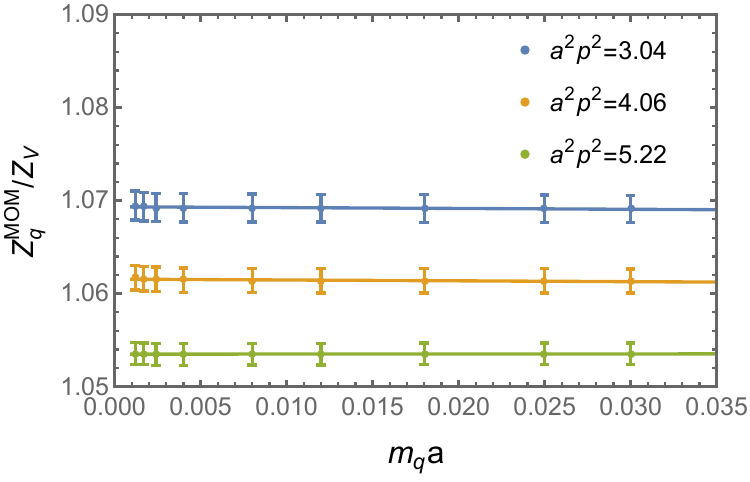}
	\includegraphics[scale=0.7]{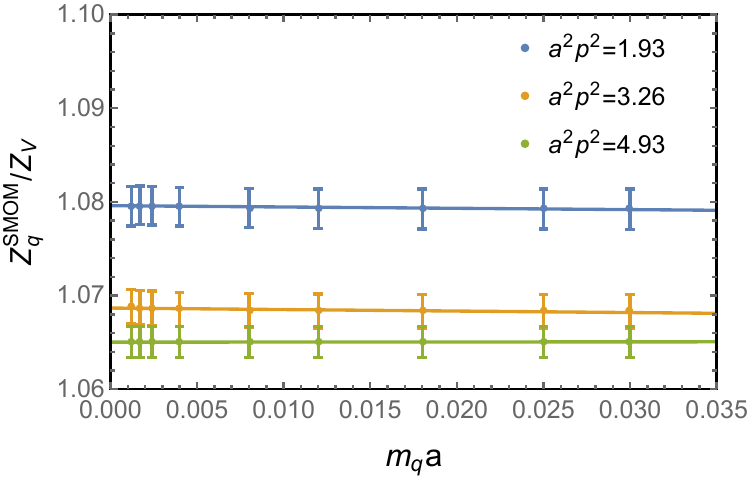}
\caption{The valence quark mass dependence of $Z^\MOM_q/Z_V$ and $Z^\SMOM_q/Z_V$ at different $a^2p^2$  on the 64I ensemble. The lines in each of the two cases are obtained by linear fits.}
\label{fig:mass_Zq}
\end{figure}

In Fig.~\ref{fig:mass_Zq}, we show the quark mass dependence of the quark field renormalization constant $Z_q^\MOM/Z_V$ (upper panel, based on Eq.~(\ref{eq:rimom_vex})) and $Z_q^\SMOM/Z_V$ (lower panel, based on Eq.~(\ref{eq:Zq_smom_vex})) on the 64I ensemble. One can see that $Z_q$ weakly depends on the quark mass in both the MOM and SMOM schemes.  As in Ref~\cite{Bi:2017ybi}, the results of the $\MOM$ and $\SMOM$ schemes in the massless limit can be obtained through linear extrapolations, 
\begin{equation}
\frac{Z_q^x}{Z_V}(m_qa)=\frac{Z_q^x}{Z_V}+c~m_qa,  
\end{equation}
where $x$ represents the $\MOM$ or $\SMOM$ scheme. $\frac{Z_q^{x}}{Z_V}$ are the RCs in the chiral limit and the solid lines in these two figures represent the central values of fits.

\begin{figure}[]
\begin{center}
	\includegraphics[scale=0.7]{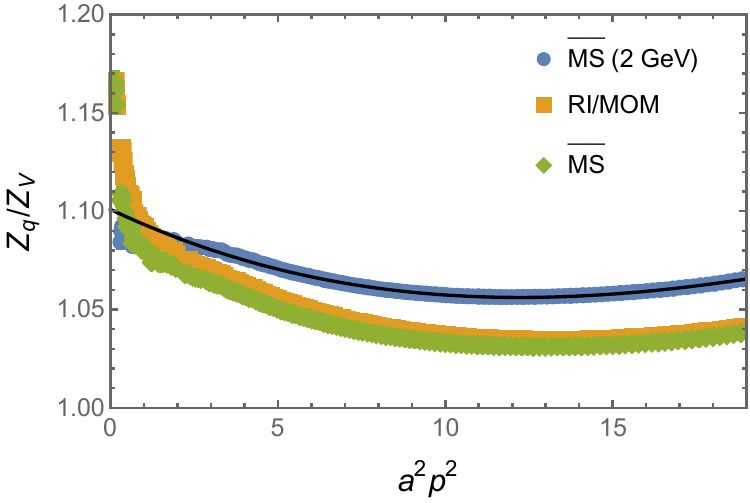}
	\includegraphics[scale=0.7]{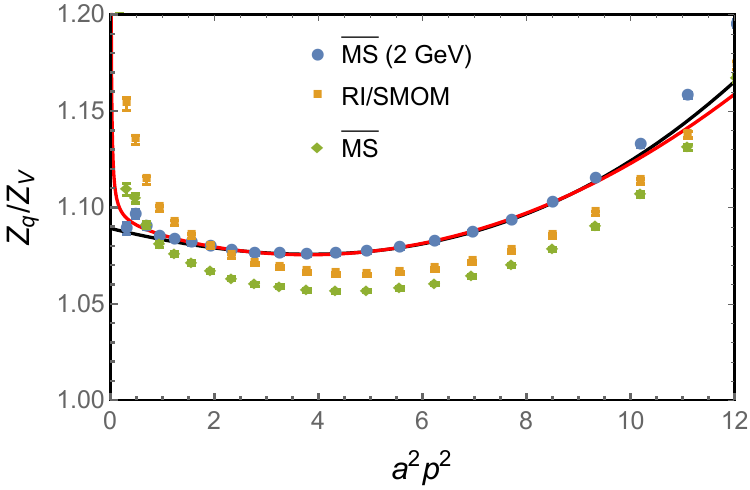}
\caption{Conversion of $Z_q/Z_V$ in the $\MOM$ (upper panel) and $\SMOM$ (lower panel) schemes to the $\MSbar$ scheme,  on the 64I ensemble. 
The yellow data represent the results in the $\MOM$ (upper panel) and $\SMOM$ (lower panel) schemes.  The green data are the results in the $\MSbar$ scheme with  $\mu=|p|$. The results in the $\MSbar$ scheme after running to 2\,GeV are shown by the blue data. The black curves represent the extrapolation of $a^2p^2$ with Eq.~(\ref{eq:fitzq_mom}) using the data in $p^2\in[9\,\GeV^2,18/a^2]$ and using Eq.~(\ref{eq:fitzq_smom1}) with the data in $a^2p^2\in[1.5,9.0]$ for the $\MOM$ and $\SMOM$ schemes, respectively. The red curve in the lower panel is from the fit using Eq.~(\ref{eq:fitzq_smom2}) with the data in $a^2p^2\in[0.3,9.0]$.}
\label{fig:Zq_momentum}
\end{center}
\end{figure}

The chiral extrapolated results can be converted to the $\MSbar$ scheme by the following matching factors~\cite{Chetyrkin:1999pq,Bi:2017ybi}
\begin{eqnarray}\label{eq:zq_convert}
&&C_q^{\MSbar,\MOM}=1+\left[-\frac{517}{18}+12\zeta_3+\frac{5}{3}n_f\right]\left(\frac{\alpha_s}{4\pi}\right)^2+   \non\\
&&\Bigg[-\frac{1287283}{648}+\frac{14197}{12}\zeta_3 +\frac{79}{4}\zeta_4
 -\frac{1165}{3}\zeta_5+\frac{18014}{81}n_f  \non\\
&&-\frac{368}{9}\zeta_3n_f-\frac{1102}{243}n_f^2\Bigg]\left(\frac{\alpha_s}{4\pi}\right)^3+\mathcal{O}(\alpha_s^4),
\end{eqnarray}

\begin{eqnarray}\label{eq:zq_convert2}
&&C_q^{\MSbar,\SMOM}=1+\left[-\frac{359}{9}+12\zeta_3+\frac{7}{3}n_f\right]\left(\frac{\alpha_s}{4\pi}\right)^2+   \non\\
 &&\Bigg[-\frac{439543}{162}+\frac{8009}{6}\zeta_3 +\frac{79}{4}\zeta_4
  -\frac{1165}{3}\zeta_5+\frac{24722}{81}n_f-   \non\\
  && \frac{440}{9}\zeta_3 n_f-\frac{1570}{243}n_f^2\Bigg]
  \left(\frac{\alpha_s}{4\pi}\right)^3+\mathcal{O}(\alpha_s^4),
\end{eqnarray}
where $\zeta_n$ is the Riemann zeta function, and $n_f=3$ is the number of light quark flavors.
The results in the intermediate scheme with $|p|$ can be converted to the $\MSbar$ scheme at $\mu=|p|$ using the above matching factors and the strong coupling constants at the same scale. To obtain the 2\,\GeV \,results in the $\MSbar$ scheme, we use the perturbative running from energy scale $\mu=|p|$ to 2\,\GeV. The anomalous dimension of the quark field $\gamma_q^\MSbar$ has been calculated up to four loops in the Landau gauge~\cite{Chetyrkin:1999pq}. With such an anomalous dimension, one can evolve the value of $Z_q^\MSbar/Z_V$ from energy scale $\mu$ to 2\,\GeV. 
The conversions of $Z_q^\MOM/Z_V$ and $Z_q^\SMOM/Z_V$ to $Z_q^\MSbar/Z_V$(2\,\GeV) for the 64I ensemble are plotted in Fig.~\ref{fig:Zq_momentum}.

The results in the upper panel of Fig.~\ref{fig:Zq_momentum} are those using the intermediate $\MOM$ scheme; 
The yellow data represent the $Z_q/Z_V$ in the $\MOM$ scheme at different $a^2p^2$ and the green data are the results in the $\MSbar$ scheme at $\mu=|p|$. The blue data are the results in the $\MSbar$ scheme with $2~\GeV$, which is evolved from $\mu$.
They exhibit nonnegligible discretization errors, especially in the large $a^2p^2$ region. The data show a downward trend in the range $3\leq a^2p^2 \leq 12$ and a gentler upward trend for $a^2p^2\geq 12$. In order to remove such a discretization error, we use following ansatz to fit the blue data,
\begin{eqnarray}\label{eq:fitzq_mom}
\frac{Z_q^\MSbar}{Z_V}(a^2p^2)&=&\frac{Z_q^\MSbar}{Z_V}+\sum_{i=1}^3C_i^{q,M}(a^2p^2)^i.
\end{eqnarray}
The fit range is chosen to be $p^2\in[9\,\GeV^2,18/a^2]$ and the $\chi^2/\text{d.o.f.}$ of the fit is less than one. We choose the same interval for the fit region for the results on the other ensembles; the lower limit of the fit region is fixed at $p^2=9\,\GeV^2$, and the upper limit is fixed at $a^2p^2=18$ since the infrared effect is dependent on $p^2$ and the discretization error in the larger $p^2$ region is sensitive to $a^2p^2$. The black curve in the upper panel of Fig.~\ref{fig:Zq_momentum} is constructed from the fit parameters, and it agrees with the lattice data. We get $Z_q^\MSbar/Z_V(2\,\GeV)$ equals 1.1009(5) after extrapolating $a^2p^2$ to zero. The fit range, fit results and $\chi^2$/d.o.f.\ for the other ensembles are listed in Table~\ref{tab:sum_FR_mom}. For the two largest lattice spacing ensembles (24D and 24DH), we apply linear extrapolations to remove the $a^2p^2$ dependence.
The fit results of the coefficients $C_i^{q,M}$ for the different ensembles are listed in Table~\ref{tab:coff_zqmom}; these discretization error terms decrease with decreasing lattice spacing.

\begin{table}[htbp]
  \centering
  \begin{tabular}{c|cccc}
  \toprule
Ensemble & $C_1^{q,M}$ & $C_2^{q,M}$ &  $C_3^{q,M}$ \\
\hline
HISQ12L & -0.01033(39) & 0.000596(37) & $-9.2(1.1)\times10^{-6}$ \\
\hline
HISQ12H & -0.01107(48) & 0.000628(46) & $-9.4(1.4)\times10^{-6}$ \\
\hline
HISQ09 & -0.00796(17) & 0.000427(18) & $-5.4(0.6)\times10^{-6}$ \\
\hline
HISQ06 & -0.00565(44) & 0.000272(50) & $-2.2(1.7)\times10^{-6}$ \\
\hline
HISQ04 & -0.00395(10) & 0.000151(12) & \ \ $0.5(0.4)\times10^{-6}$ \\
\hline
48I & -0.01059(33) & 0.000592(32) & $-8.5(1.0)\times10^{-6}$ \\
\hline
64I & -0.00806(17) & 0.000420(18)  & $-4.9(0.6)\times10^{-6}$ \\
\hline
48If & -0.00629(10) & 0.000293(12) & $-2.1(0.4)\times10^{-6}$ \\
\hline
32If & -0.00456(17) & 0.000162(21) & \ \ $1.1(0.8)\times10^{-6}$ \\
\hline
\hline
  \end{tabular}
\caption{Fit results of the coefficients for the $\MOM$ scheme; the corresponding fit ansatz is Eq.~(\ref{eq:fitzq_mom}). }. 
  \label{tab:coff_zqmom}
\end{table}

In addition to the statistical errors, there are a series of systematic errors that need to be considered:

1) Conversion ratio between the $\MOM$ and $\MSbar$ schemes: When $n_f=3$, the matching factor  in Eq.~(\ref{eq:zq_convert}) can be written as  
\begin{eqnarray}\label{eq:zq_trun}
C_q^{\MSbar,\MOM}(\mu^2)&=&1-0.0589\alpha_s^2(\mu^2)-0.2352\alpha_s^3(\mu^2)\nonumber\\
&&+\mathcal{O}(\alpha_s^4), 
\end{eqnarray}
assuming the coefficient of the $\mathcal{O}(\alpha_s^4)$ term is 4 ($\approx$ 0.2352/0.0589) times larger than that of the $\mathcal{O}(\alpha_s^3)$ term, we find that the central value of $Z_q^\MSbar/Z_V$(2$\,\GeV$) becomes 1.0971 if we include  such a dummy $\mathcal{O}(\alpha_s^4)$ term in the matching factor for each of the $\mu^2=p^2$; thus we estimate the systematic error from the conversion ratio by the difference of the central values, which is $(1.1009-1.0971)/1.1009=0.34\%$. Such a estimation is different from our previous strategy~\cite{Bi:2017ybi}, where we  chose the correction of this dummy $\mathcal{O}(\alpha_s^4)$ term only at the smallest $p^2$ used in the fit to estimate the systematic error.

2) Perturbative running: The $\MSbar$ result at 2\,GeV has been obtained with the quark field anomalous dimension up to four loops \cite{Chetyrkin:1999pq}; the systematic error caused by the anomalous dimension can be estimated through using the anomalous dimension up to three loops to do the perturbative running. The error caused by perturbative running is about 0.03$\%$.

3) $\Lambda^\MSbar_\QCD$: In our calculation, $\Lambda^\MSbar_\QCD$ is chosen to be 0.332(17)\,GeV for three flavors \cite{ParticleDataGroup:2016lqr}. The 1$\sigma$ deviation of $\Lambda^\MSbar_\QCD$ will cause the central value of $Z_q^\MSbar/Z_V$(2$\,\GeV$) to shift by 0.02$\%$. 

4) Lattice spacing: The lattice spacing of the 64I ensemble, 0.0837(2)\,fm, has its uncertainty. If we modify the lattice spacing by 1$\sigma$ and redo all the procedures, the $Z_q$ we get will differ by 0.01\% which should be considered as a systematic uncertainty.

5) Fit range of $a^2p^2$: If using $p^2\in[9\,\GeV^2,15/a^2]$ to do the $a^2p^2$ extrapolation, the central value of $Z_q^\MSbar/Z_V$(2$\,\GeV$) becomes 1.0995 and the uncertainty caused by the change of the fit range is about 0.13$\%$. 

6) Finite volume effect: {In order to estimate the finite volume effect, we dropped the lightest two quark masses in the chiral extrapolation, which have relatively larger finite volume effects. It introduces a 0.02\% change on $Z_q$.}

\begin{table}[htbp]
  \centering
  \begin{tabular}{c|cccc}
  \toprule
$m_la$ & 0.005 & 0.01 & 0.02 & 0.03  \\
\hline
$Z_q^\MSbar/Z_V(2\,\GeV)$ & 1.1035(1) & 1.1041(1) & 1.1054(1) & 1.1061(1) \\
\hline
$\tilde{Z}_q^\MSbar/Z_V(4\,\GeV)$ & 1.0848(1) & 1.0852(1) & 1.0863(1) & 1.0872(1) \\
\hline
  \end{tabular}
\caption{The results of $Z_q^\MSbar/Z_V(2\,\GeV)$ and $\tilde{Z}_q^\MSbar/Z_V(4\,\GeV)$ on the 24I ensemble from the intermediate $\MOM$ and $\SMOM$ schemes, respectively. The results from the $\MOM$ scheme have been extrapolated to the $a^2p^2\rightarrow 0$ limit, but the $\tilde{Z}_q^\MSbar/Z_V(4\,\GeV)$ are obtained from the $\SMOM$ scheme at the specific $a^2p^2=4.93$ where $|p|\sim 4$\,GeV. The corresponding slopes from the linear extrapolations of the light sea quark are 0.038(2)$\,\GeV^{-1}$ and 0.035(1)$\,\GeV^{-1}$, respectively.}
  \label{tab:sea_dep_Zq}
\end{table}

\begin{table}[htbp]
  \centering
  \begin{tabular}{ccccc}
  \toprule
Source &  $Z_q^\MSbar/Z_V$ &$Z_S^\MSbar/Z_V$ &$Z_P^\MSbar/Z_V$ &$Z_T^\MSbar/Z_V$ \\
\hline
Statistical  & 0.04  & 0.08 & 0.21 & 0.01\\
Conversion ratio   & 0.34  & 2.29 & 2.15 & 0.40\\
Perturbative running & 0.03 & 0.11 & 0.11 & 0.03\\
$\Lambda_{\rm QCD}^{\overline{\text{MS}}}$ & 0.02 & 0.31 & 0.26 & 0.04\\ 
Lattice spacing & 0.01 & 0.09 & 0.09 & 0.03\\
Fit range of $a^2p^2$   & 0.13 & 0.03 & 0.27 & 0.01 \\
Finite-volume effect   & 0.02 & 0.07 & 0.14 & 0.01 \\
$m_s^{\text{sea}}\neq$ 0 &  0.17 & 0.46 & 1.61 & 0.06\\
Total uncertainty  & 0.41 & 2.36 & 2.73 & 0.41\\
\hline
  \end{tabular}
  \caption{Summary of uncertainties of RCs in percentage on the 64I ensemble through the intermediate $\MOM$ scheme.}
  \label{tab:sum_error}
\end{table}

7) Nonzero sea strange quark mass: In order to estimate the effect of nonzero strange quark mass, we calculated $Z_q^\MSbar/Z_V(2\,\GeV)$ on the four 24I ensembles and the results are listed in Table~\ref{tab:sea_dep_Zq}. With the linear light sea quark mass extrapolation we find the slope is about 0.038(2)\,$\GeV^{-1}$, from which we find an error of 0.17\% due to the nonzero strange quark mass. 

The summary for the uncertainties of $Z_q^\MSbar/Z_V$ is presented in Table~\ref{tab:sum_error}. The final result of $Z_q^\MSbar/Z_V$ on the 64I ensemble equals 1.1009(45), from which we obtain 
$Z_q^\MSbar(2\,\GeV)=1.188(5)$ given $Z_V=Z_A=1.0787(1)(1)$ from the AWI. 


In the SMOM case, we also choose the linear chiral extrapolation model to obtain the $Z_q^\SMOM/Z_V$ in the chiral limit. After converting $Z_q^\SMOM/Z_V$ to the $\MSbar$ scheme and running the energy scale to 2\,GeV, we obtain the results of $Z_q^\MSbar/Z_V(2\,\GeV)$ again, which are presented in the lower panel of Fig.~\ref{fig:Zq_momentum}. Comparing the results from the $\MOM$ and $\SMOM$ schemes, 
we can see that the results from the $\SMOM$ scheme have stronger nonlinear dependence on $a^2p^2$. We use the following ansatzes to fit the blue data in the lower panel,
\begin{eqnarray}\label{eq:fitzq_smom1}
\frac{Z^\MSbar_q}{Z_V}(a^2p^2)&=&\frac{Z^\MSbar_q}{Z_V}+\sum_{i=1}^3C_i^{q,S}(a^2p^2)^i
\end{eqnarray}
and
\begin{eqnarray}\label{eq:fitzq_smom2}
\frac{Z^\MSbar_q}{Z_V}(a^2p^2)&=&\frac{Z^\MSbar_q}{Z_V}+\frac{C^{q,S}_{-1}}{a^2p^2}+\sum_{i=1}^3C_i^{q,S}(a^2p^2)^i, 
\end{eqnarray}
where the pole term in Eq.~(\ref{eq:fitzq_smom2}) occurs in the operator product expansion of the quark propagator~\cite{Pascual:1981jr}. The fit results and $\chi^2$/d.o.f.\ are listed in Table~(\ref{tab:zq_ap_dep}).
The fit results are quite sensitive to the fit region, due to the 
highly nonlinear $a^2p^2$ dependence.
Finally, we take fit result corresponding to $a^2p^2\in[1.5,9.0]$ as the central value and statistical error. The deviation between the results obtained by fitting the data in the range $a^2p^2\in[1.5,9.0]$ and $a^2p^2\in[3.0,10.5]$ is about $1.01\%$, which is much larger than that in the RI/MOM case and we use this deviation to estimate the systematic error due to the fit range. We also take the discrepancy between the fit results of Eq.~(\ref{eq:fitzq_smom1}) and Eq.~(\ref{eq:fitzq_smom2}) to be the systematic error, which is about 0.23$\%$. One can also use Eq.~(\ref{eq:fitzq_smom2}) to fit the $Z^\MSbar_q$ data through the RI/MOM scheme, wherein the coefficient of the $1/(a^2p^2)$ term is found to be consistent with zero so that the result is unchanged.

\begin{table}[htbp]
  \centering
  \begin{tabular}{c|c|cc}
  \toprule
Fit ansatz & Fit Range for $a^2p^2$ & Result & $\chi^2$/d.o.f. \\
\hline
\multirow{3}{*}{Eq.(\ref{eq:fitzq_smom1})} 
&[1.0,8.0] & 1.0931(30) & 0.09 \\
&[1.5,9.0] & 1.0891(46) & 0.10\\
&[3.0,10.5] & 1.0781(94) & 0.07 \\
\hline
\multirow{1}{*}{Eq.(\ref{eq:fitzq_smom2})} 
&[0.3,9.0] & 1.0916(37) & 0.72 \\
\hline
\hline
  \end{tabular}
\caption{The fit results of $Z_q^\MSbar/Z_V$(2$\,\GeV$) {through the intermediate $\SMOM$ scheme  on the 64I ensemble}, for different fit ranges of $a^2p^2$.}
  \label{tab:zq_ap_dep}
\end{table}

When $n_f=3$, the matching factor between the $\SMOM$ scheme and the $\MSbar$ scheme can be rewritten as
\begin{eqnarray}
C_{q}^{\MSbar,\SMOM}&=&1-0.1169\alpha_s^2-0.4076\alpha_s^3+\mathcal{O}(\alpha_s^4).
\end{eqnarray}
Since both ${\cal O}(\alpha_s^2)$ and ${\cal O}(\alpha_s^3)$ are larger than in the MOM case, the conversion ratio here will introduce a larger systematic uncertainty. We also estimate the other systematic uncertainties with similar strategies to the $\MOM$ case, except that of the nonzero strange quark mass. Since the available data points in the SMOM case on the 24I ensemble are not as many as those on the ensembles with larger volume, we just estimate the uncertainty caused by the nonzero strange quark mass by the results $\tilde{Z}_q^\MSbar/Z_V$ from the $\SMOM$ at $a^2p^2=4.93$, as the effect of nonzero sea quark mass should not be very sensitive to $a^2p^2$. Here we use the $\tilde{Z}_q^\MSbar$ to represent the results from the $\SMOM$ scheme to distinguish the result from $\MOM$ scheme, where we calculate the results at different $a^2p^2$ then extrapolate to the $a^2p^2\rightarrow 0$ limit. We list the results $\tilde{Z}_q^\MSbar/Z_V$ on the 24I ensembles in the third row in Table~\ref{tab:sea_dep_Zq}. 

{Eventually we get $Z_q^\MSbar(2\,\GeV)$ on the 64I ensemble to be 1.175(16) through the intermediate $\SMOM$ scheme. It is consistent with that through the RI/MOM scheme except it has a factor of 3 larger uncertainty majorly from the fit range.}
 
\begin{table}[htbp]
  \centering
  \begin{tabular}{ccccc}
  \toprule
Source &  $Z_q^\MSbar/Z_V$ &$Z_S^\MSbar/Z_V$ &$Z_P^\MSbar/Z_V$ &$Z_T^\MSbar/Z_V$ \\
\hline
Statistical  & 0.42  & 0.59 & 0.63 & 0.23\\
Conversion ratio   & 0.75  & 0.23 & 0.22 & 0.81\\
Perturbative running & 0.01 & 0.09 & 0.08 & 0.01\\
$\Lambda_{\rm QCD}^\MSbar$ & 0.07 & 0.18 & 0.18 & 0.04\\ 
Lattice spacing & 0.01 & 0.06 & 0.07 & 0.04\\
Fit range of $a^2p^2$   & 1.01 & 1.29 & 1.79 & 0.36 \\
Finite-volume effect   & 0.01 & 0.02 & 0.32 & 0.01 \\
$m_s^{\text{sea}}\neq$ 0 &  {0.15} & {0.05} & {0.05} & {0.12}\\
Different fit models    & 0.23  & 5.10 & 5.20 & 0.46 \\
Total uncertainty  & 1.36 & 5.30 & 5.55 & 1.03\\
\hline
  \end{tabular}
  \caption{Summary of uncertainties of RCs in percentage on the 64I ensemble through the intermediate $\SMOM$ scheme.}
  \label{tab:sum_error_S}
\end{table}

\begin{figure*}[]
\begin{center}
\subfigure[]
{
	\begin{minipage}[b]{0.45\linewidth}
	\centering 
	\includegraphics[scale=0.65]{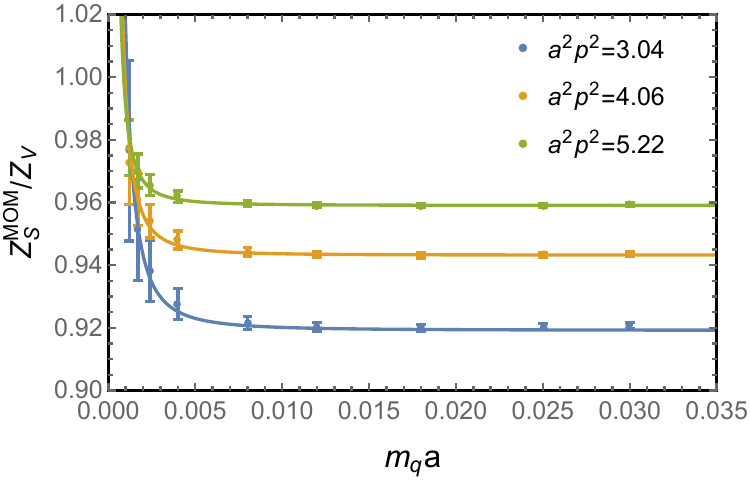}
	\end{minipage}
}
\subfigure[]
{
	\begin{minipage}[b]{0.45\linewidth}
	\centering  
	\includegraphics[scale=0.65]{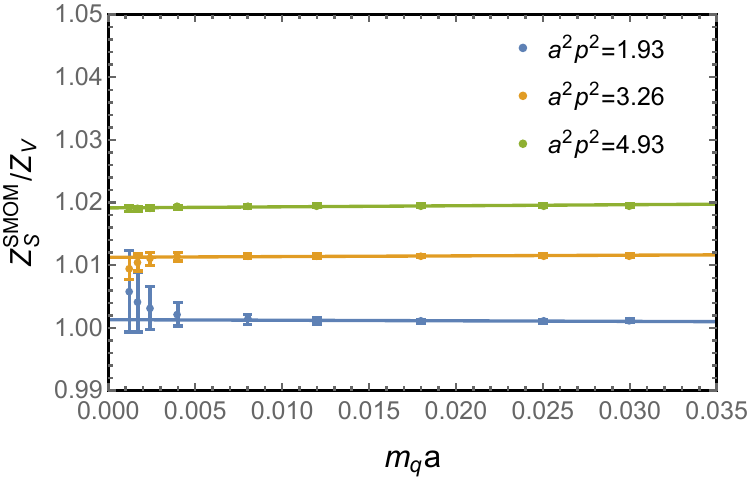}
	\end{minipage}
}
\caption{The valence quark mass dependence of $Z^\MOM_S/Z_V$ and $Z^\SMOM_S/Z_V$  on the 64I ensemble with different $a^2p^2$. The curves are the fits using the ansatzes in  Eq.~(\ref{eq:ZS_massfit}) and Eq.~(\ref{eq:chiral_extra_smom}) for the $\MOM$ and $\SMOM$ schemes.}
\label{fig:mass_Zs}
\end{center}
\end{figure*}

\begin{figure*}[]
\begin{center}
\subfigure[]
{
	\begin{minipage}[b]{0.45\linewidth}
	\centering 
	\includegraphics[scale=0.65]{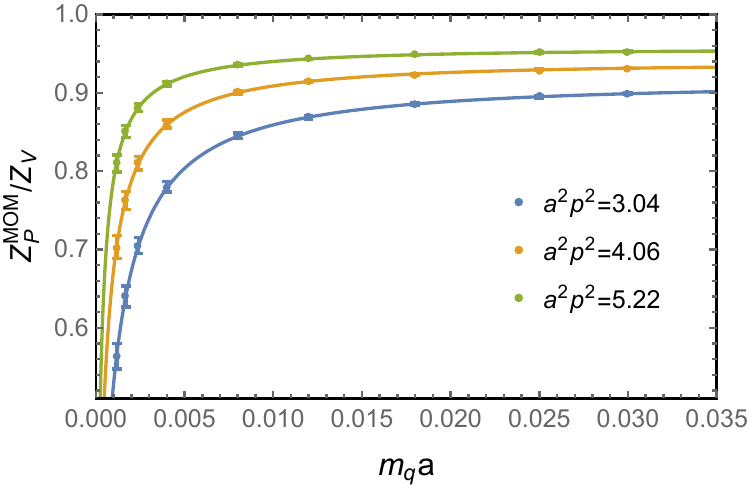}
	\end{minipage}
}
\subfigure[]
{
	\begin{minipage}[b]{0.45\linewidth}
	\centering  
	\includegraphics[scale=0.65]{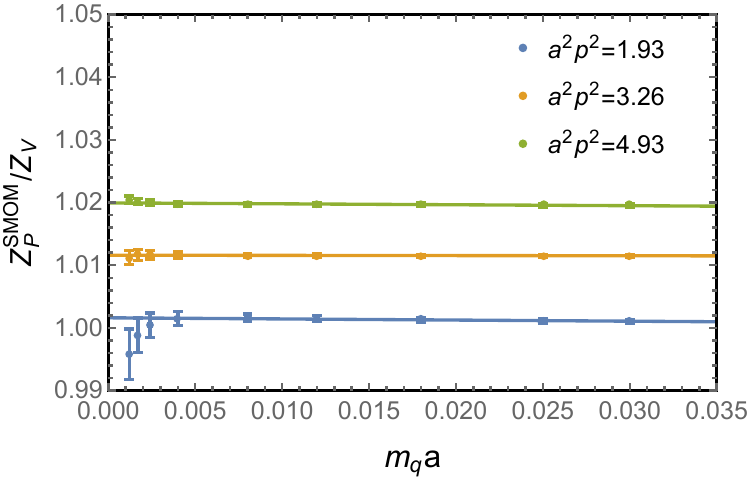}
	\end{minipage}
}
\caption{The valence quark mass dependence of $Z^\MOM_P/Z_V$ and $Z^\SMOM_P/Z_V$  on the 64I ensemble at different $a^2p^2$. The curves in the left panel are the fits using Eq.~(\ref{eq:ZP_massfit}). The curves in the right panel are obtained by linear extrapolation.}
\label{fig:mass_Zp}
\end{center}
\end{figure*}

\subsection{Scalar and pseudoscalar currents renormalization}\label{sec:s_and_ps}

{We start from} the unphysical mass pole of the renormalization constants of the scalar operator and pseudoscalar operator.
The renormalization constants of the scalar and pseudoscalar operators in the $\MOM$ and $\SMOM$ schemes can be obtained through
\begin{eqnarray}
\frac{Z_{S,P}^{\MOM}}{Z_V^{\MOM}}&=&\frac{\Gamma_V(p)}{\Gamma_{S,P}(p)}\bigg|_{p^2=\mu^2},  \non \\
\frac{Z_{S,P}^{\SMOM}}{Z_V^{\SMOM}}&=&\frac{\Gamma_V(p_1,p_2)}{\Gamma_{S,P}(p_1,p_2)}\bigg|_{sym},
\end{eqnarray}
where 
\begin{eqnarray}
&&\Gamma_{S}(p)=\frac{1}{12}\Tr[\Lambda_{S,B}(p)I],~
\Gamma_{P}(p)=\frac{1}{12}\Tr[\Lambda_{P,B}(p)\gamma_5]  ,\non\\
&&\Gamma_{S}(p_1,p_2)=\frac{1}{12}\Tr[\Lambda_{S,B}(p_1,p_2)I],\non\\
&&\Gamma_{P}(p_1,p_2)=\frac{1}{12}\Tr[\Lambda_{P,B}(p_1,p_2)\gamma_5],
\end{eqnarray}
and 
\begin{eqnarray}
&&\Gamma_{V}(p)=\frac{1}{48}\Tr[\Lambda^\mu_{V,B}(p)\gamma_\mu],\non\\
&&\Gamma_{V}(p_1,p_2)=\frac{1}{12q^2}\Tr[q_\mu\Lambda^\mu_{V,B}(p_1,p_2)\slashed{q}].
\end{eqnarray}
The results of $Z_S^\MOM/Z_V$ on the 64I ensemble versus the valence quark mass when $a^2p^2$=3.04, 4.06 and 5.22 are presented in the left panel of Fig.~\ref{fig:mass_Zs}. One can see that the value of $Z_S^\MOM/Z_V$ diverges when the quark mass decreases toward zero, especially when $a^2p^2$ is small. The reason for the mass pole is that the amputated Green function of the scalar quark operator obtains a large contribution from zero modes of the Dirac operator in the chiral limit~\cite{Blum:2001sr}, and it causes the renormalization constants $Z_S$ to have a power divergence in the valence quark mass $m_q$. To extract the RCs in the chiral limit, we use the following ansatz to fit the lattice results in the RI/MOM scheme as in Ref.~\cite{Blum:2001sr,Aoki:2007xm, Liu:2013yxz,Bi:2017ybi},
\begin{equation}\label{eq:ZS_massfit}
\frac{Z_S^\MOM}{Z_V}(am_q)=\frac{A_s}{(am_q)^2}+B_s+C_sam_q,
\end{equation}
where $B_s$ is the result of $Z_S^\MOM/Z_A$ in the chiral limit. The $\chi^2$/d.o.f.\ of the fit is around 1 and the curves in Fig.~\ref{fig:mass_Zs} are constructed using the fit parameters and they agree with the data. 

For $Z_P/Z_V$, the unphysical pole in the $\MOM$ scheme is inversely proportional to $1/am_q$ or $1/m_\pi^2$, which corresponds to the mass pole of the Goldstone meson~\cite{Becirevic:2004ny}. The quark mass dependence of $Z_P^\MOM/Z_V$ when $a^2p^2=3.04$, 4.06, 5.22 is shown in {the left panel of} Fig.~\ref{fig:mass_Zp}. One can see that $Z_P^\MOM/Z_V$ will approach zero in the chiral limit. To subtract the contamination of the unphysical quark mass pole, we use the following ansatz to fit the lattice data,
\begin{equation}\label{eq:ZP_massfit}
\frac{Z_P^\MOM}{Z_V}(am_q)=\frac{1}{\frac{A_p}{(am_q)}+B_p+C_pam_q},  
\end{equation}
where $1/B_p$ is the result of $Z_P^\MOM/Z_V$ in the chiral limit. {The curves show that the fit predictions are consistent with the data}.

We plot the mass dependence of {$Z_{S/P}^\SMOM/Z_V$ in the right panels of Fig.~\ref{fig:mass_Zs} and ~\ref{fig:mass_Zp}, respectively.} Comparing with the results of the $\MOM$ scheme, one can see that the quark mass dependence of $Z_{S/P}^\SMOM/Z_V$ is {free of the unphysical pole}.  
So we use the following ansatz to extrapolate the results to the chiral limit, 
\begin{equation}\label{eq:chiral_extra_smom}
\frac{Z_{S/P}^\SMOM}{Z_V}(am_q)=\tilde{B}_{s/p}+\tilde{C}_{s/p}am_q,
\end{equation}
and the fit results {agree with the data well.}

\begin{figure*}[]
\begin{center}
\subfigure[]
{
	\begin{minipage}[b]{0.45\linewidth}
	\centering 
	\includegraphics[scale=0.65]{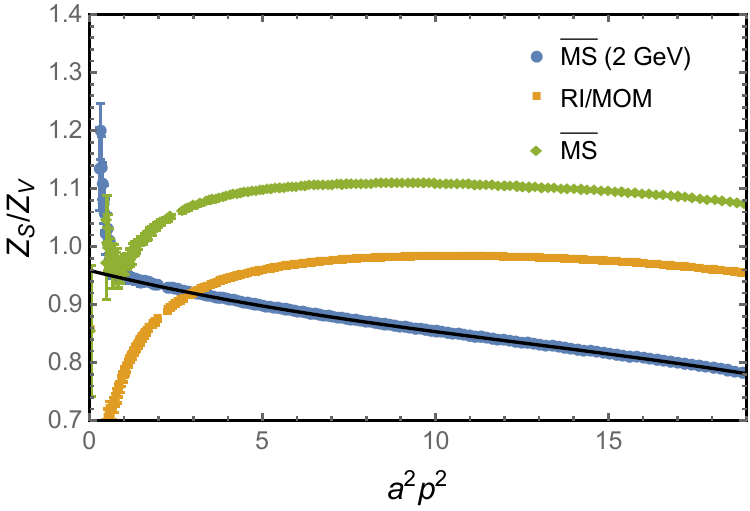}
	\end{minipage}
}
\subfigure[]
{
	\begin{minipage}[b]{0.45\linewidth}
	\centering  
	\includegraphics[scale=0.65]{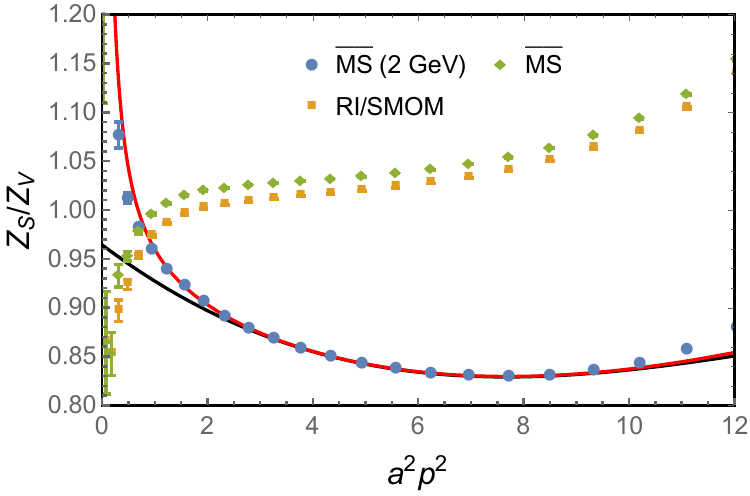}
	\end{minipage}
}
\caption{Conversion and running of $Z_S^\MSbar/Z_V$ from the $\MOM$ scheme (left panel) and $\SMOM$ scheme (right panel)  on the 64I ensemble. The yellow data represent the results in the $\MOM$ scheme (left panel) and $\SMOM$ scheme (right panel).  The green data are the results in the $\MSbar$ scheme with  $\mu=|p|$. The results in the $\MSbar$ scheme after running to 2\,GeV are shown by the blue data.
The black curves represent the fit using Eq.~(\ref{eq:fitZs_mom}) with data in $p^2\in[9\,\GeV^2,18/a^2]$ and Eq.~(\ref{eq:fitZs_mom}) with data in $a^2p^2\in[3.5,9]$ for the $\MOM$ and $\SMOM$ schemes, respectively. The red curve in the right panel is the fit using Eq.~(\ref{eq:ZS_extrapo2}) with data in $a^2p^2\in[1,9]$.}
\label{fig:Zs_momentum}
\end{center}
\end{figure*}

After the chiral extrapolation, the result in the $\MSbar$ scheme can be obtained by using the following matching factor~\cite{Franco:1998bm,Chetyrkin:1999pq}
\begin{eqnarray}\label{eq:zs_convert}
&&C_S^{\MSbar,\MOM}=\frac{Z_S^\MSbar/Z_V^\MSbar}{Z_S^\MOM/Z_V^\MOM}=\frac{Z_m^\MOM}{Z_m^\MSbar}=  \non\\
&&1+\frac{16}{3}\left(\frac{\alpha_s}{4\pi}\right)+\left(\frac{2246}{9} -\frac{89n_f}{9}-\frac{152\zeta_3}{3}\right)\left(\frac{\alpha_s}{4\pi}\right)^2    \non\\
&&+\Bigg(\frac{8290535}{648}-\frac{262282n_f}{243}   +\frac{8918n_f^2}{729}-\frac{466375\zeta_3}{108}     \non\\
&&+\frac{4936\zeta_3n_f}{27}+\frac{32\zeta_3n_f^2}{27}-\frac{80\zeta_4n_f}{3}+\frac{2960\zeta_5}{9}\Bigg)\left(\frac{\alpha_s}{4\pi}\right)^3
\non\\
&&+\mathcal{O}(\alpha_s^4).
\end{eqnarray}
We have used the relation $Z_V^\MSbar=Z_V^{\MOM}=Z_V^{\SMOM}$ since $Z_V=Z_A$ and the latter is determined by the PCAC relation and it should be independent of the renormalization scheme. The anomalous dimension of the scalar operator can be obtained through the relation $\gamma_{\bar{\psi}\psi}=-\gamma_m$, where $\gamma_m$ is the anomalous dimension of quark mass and it has been calculated up to three loops in Ref.~\cite{Chetyrkin:1999pq}. The result of $Z_S^\MSbar/Z_V$(2\,$\GeV$) from the intermediate $\MOM$ scheme 
is plotted in the left panel of Fig.~\ref{fig:Zs_momentum}. 
The yellow data are the results of $Z_S/Z_V$ in the $\MOM$ scheme at different $a^2p^2$ and the green data are the results in the $\MSbar$ scheme at $\mu=|p|$. The blue data are the results in the $\MSbar$ scheme with $2~\GeV$, which is evolved from $\mu$.
We use the following ansatz to fit the blue data
\begin{eqnarray}\label{eq:fitZs_mom}
\frac{Z_S^\MSbar}{Z_V}(a^2p^2)&=&\frac{Z_S^\MSbar}{Z_V}+\sum_{i=1}^3C^{S,M}_i(a^2p^2)^i,
\end{eqnarray}
where the fit range is $p^2\in[9\,\GeV^2,18/a^2]$ and the fit result is $Z_S^\MSbar/Z_V(2\,\GeV)=0.9588(8)$. The solid line in Fig.~\ref{fig:Zs_momentum} is the fit using Eq.~(\ref{eq:fitZs_mom}). The fit range, fit results and $\chi^2$/d.o.f.\ for the other ensembles are listed in Table~\ref{tab:sum_FR_mom}. The fit results of the coefficients $C_i^{S,M}$ for the different ensembles are listed in Table~\ref{tab:coff_zsmom}; the absolute values of these coefficients decrease with decreasing lattice spacing.

\begin{table}[htbp]
  \centering
  \begin{tabular}{c|cccc}
  \toprule
Ensemble & $C_1^{S,M}$ & $C_2^{S,M}$ &  $C_3^{S,M}$ \\
\hline
HISQ12L & $-0.0283 (13)$ & 0.00134(11) & $-0.000032 (03)$ \\
\hline
HISQ12H & $-0.0254(25)$ & 0.00116(20) & $-0.000029(05)$ \\
\hline
HISQ09 & $-0.0163(05)$ & 0.00064(05) & $-0.000016(02)$ \\
\hline
HISQ06 & $-0.0128(06)$ & 0.00051(07) & $-0.000013(02)$ \\
\hline
HISQ04 & $-0.0126(02)$ & 0.00056(02) & $-0.000014(01)$ \\
\hline
48I & $-0.0166(06)$ & 0.00059(05) & $-0.000016(02)$ \\
\hline
64I & $-0.0147(02)$ & 0.00056(02)   & $-0.000015(01)$ \\
\hline
48If & $-0.0134(02)$ & 0.00050(02) & $-0.000013(01)$ \\
\hline
32If & $-0.0129(03)$ & 0.00048(03) & $-0.000012(01)$ \\
\hline
\hline
  \end{tabular}
\caption{The fit results of coefficients for the $\MOM$ scheme; the corresponding fit ansatz is Eq.~(\ref{eq:fitZs_mom}).}
  \label{tab:coff_zsmom}
\end{table}

\begin{table}[htbp]
  \centering
  \begin{tabular}{c|cccc}
  \toprule
Ensemble & $C_1^{P,M}$ & $C_2^{P,M}$ &  $C_3^{P,M}$ \\
\hline
HISQ12L & $-0.0372 (30)$ & 0.00193(24) & $-0.000045 (06)$ \\
\hline
HISQ12H & $-0.0378(75)$ & 0.00201(60)  & $-0.000049(16)$ \\
\hline
HISQ09 & $-0.0199(12)$ & 0.00091(11) & $-0.000023(03)$ \\
\hline
HISQ06 & $-0.0137(13)$ & 0.00058(13) & $-0.000015(04)$ \\
\hline
HISQ04 & $-0.0133(36)$ & 0.00062(04) & $-0.000016(01)$ \\
\hline
48I & $-0.0200(14) $ & 0.00083(12) & $-0.000022(03)$ \\
\hline
64I & $-0.0166(06)$ & 0.00071(05)   & $-0.000019(02)$ \\
\hline
48If & $-0.0170(04)$ & 0.00078(04) & $-0.000020(01)$ \\
\hline
32If & $-0.0151(07)$ & 0.00067(07)  & $-0.000017(02)$ \\
\hline
\hline
  \end{tabular}
\caption{The fit results of coefficients for the $\MOM$ scheme; the corresponding fit ansatz is Eq.~(\ref{eq:fitZp_mom}).}
  \label{tab:coff_zpmom}
\end{table}

\begin{table}[htbp]
  \centering
  \begin{tabular}{c|ccccc}
  \toprule
$m_la$ & 0.005 & 0.01 & 0.02 & 0.03  \\
\hline
$Z_S^\MSbar/Z_V(2\,\GeV)$ & 1.0083(07) & 1.0099(08) & 1.0134(07) & 1.0130(12)\\
\hline
$Z_P^\MSbar/Z_V(2\,\GeV)$ & 1.0225(23)& 1.0300(27)& 1.0361(27)& 1.0450(40) \\
\hline
\hline
$\tilde{Z}_S^\MSbar/Z_V(4\,\GeV)$ & 0.8907(1) & 0.8908(1) & 0.8913(1) & 0.8914(1)\\
\hline
$\tilde{Z}_P^\MSbar/Z_V(4\,\GeV)$ & 0.8905(1)& 0.8907(1)& 0.8910(1)& 0.8912(1) \\
\hline
\hline
  \end{tabular}
\caption{The results of $Z_S/Z_V$ and $Z_P/Z_V$ through the $\MOM$ and $\SMOM$ schemes on the 24I ensemble. The results at 2\,GeV are calculated using the intermediate $\MOM$ scheme, and the corresponding slopes from linear extrapolation of the light sea quark are 0.087(17)\,$\GeV^{-1}$ and 0.308(58)\,$\GeV^{-1}$, respectively. 
The results in the fourth and fifth rows are from the $\SMOM$ scheme at $a^2p^2=4.93$, and the slopes of $Z_S/Z_V$ and $Z_P/Z_V$ are 0.011(1)\,$\GeV^{-1}$ and 0.010(2)\,$\GeV^{-1}$, respectively.}
  \label{tab:sea_dep_ZS}
\end{table}

As in the calculation of the quark field renormalization constant $Z_q$, we also estimate the systematic error due to the truncation error of the matching factor $C_S^\SMOM$. When $n_f=3$, the matching factor in Eq.~(\ref{eq:zs_convert})
can be rewritten as
\begin{eqnarray}
C_{S, n_f=3}^{\MSbar,\MOM}&=&1+0.4244\alpha_s+1.0068\alpha_s^2 +2.7221\alpha_s^3\nonumber\\
&&+\mathcal{O}(\alpha_s^4). 
\end{eqnarray}
Similar to what we did with the renormalization of the quark field strength $Z_q$, here we also add a dummy ${\cal O}(\alpha_s^4)$ term with coefficient $2.7221^2/1.0068=7.3597$ for each of the $\mu^2=p^2$ used in the $a^2p^2$ extrapolation; we find the systematic uncertainty from the conversion ratio to be 2.29$\%$, which is more conservative than our previous estimate $1.5\%$~\cite{Bi:2017ybi}, which corresponds to the correction of ${\cal O}(\alpha_s^4)$ only at the smallest $p^2$ used in the fit. The final result of $Z_S^\MSbar/Z_V(2\,\GeV)$ on the 64I ensemble through the MOM scheme is 0.959(1)(22)(6), where the first error is statistical and the latter two uncertainties are from the conversion ratio and other systematic uncertainties.

\begin{figure*}[]
\begin{center}
\subfigure[]
{
	\begin{minipage}[b]{0.45\linewidth}
	\centering 
	\includegraphics[scale=0.65]{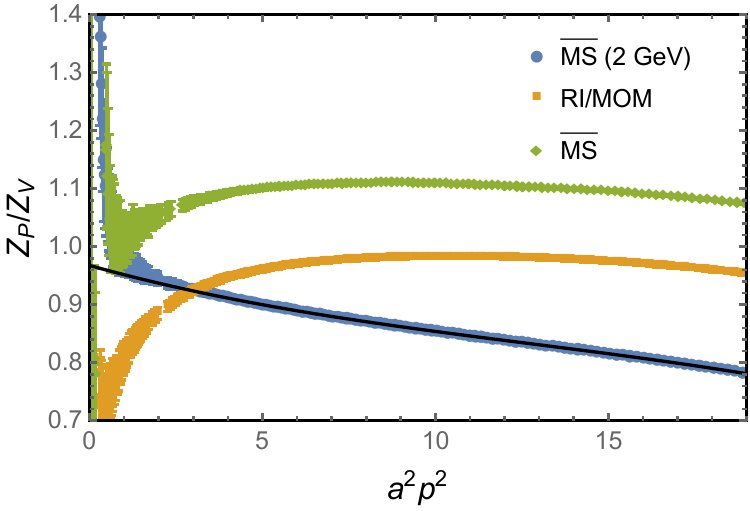}
	\end{minipage}
}
\subfigure[]
{
	\begin{minipage}[b]{0.45\linewidth}
	\centering  
	\includegraphics[scale=0.65]{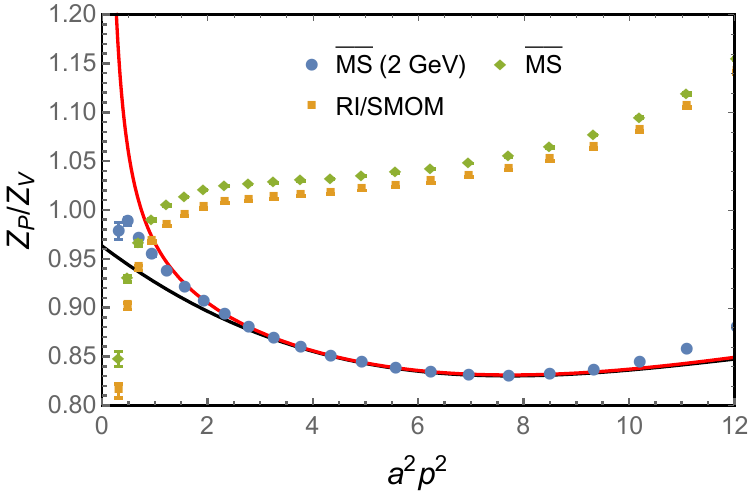}
	\end{minipage}
}
\caption{Conversion and running of $Z_P^\MSbar/Z_V$ from the $\MOM$ scheme (left panel) and $\SMOM$ scheme (right panel) on the 64I ensemble. The yellow data represent the results in the $\MOM$ scheme (left panel) and $\SMOM$ scheme (right panel).  The green data are the results in the $\MSbar$ scheme with  $\mu=|p|$. The results in the $\MSbar$ scheme after running to 2\,GeV are shown by the blue data. The black curves represent the fit using Eq.~(\ref{eq:fitZs_mom}) with data in $p^2\in[9\,\GeV^2,18/a^2]$ and Eq.~(\ref{eq:fitZs_mom}) with data in $a^2p^2\in[3.5,9]$ for the $\MOM$ and $\SMOM$ schemes, respectively. The red curve in the right panel is the fit using Eq.~(\ref{eq:ZS_extrapo2}) with data in $a^2p^2\in[1,9]$.}
\label{fig:Zp_momentum}
\end{center}
\end{figure*}

For the pseudoscalar current, the numerical results in Fig.~\ref{fig:mass_Zp} show {that its} matrix element has a pole in the chiral limit in the $\MOM$ scheme, but this pole effect is much smaller or nonexistent in the $\SMOM$ scheme. We subtract the Goldstone mass pole by using Eq.~(\ref{eq:ZP_massfit}) to fit the lattice data of $\MOM$, convert the result of $Z_P^\MOM/Z_V^\MOM$ into $Z_P^\MSbar/Z_V^\MSbar$, and then use the anomalous dimension $\gamma_S^\MSbar$ to evolve the energy scale to 2\,GeV. Since overlap fermions are chiral, the matching coefficient and anomalous dimension of the pseudoscalar operator are the same as those of the scalar operator. The results of $Z_P^\MSbar/Z_V$ from the $\MOM$ scheme are presented in the left panel of Fig.~\ref{fig:Zp_momentum}.  We use the following ansatz to fit the lattice data of $Z_P^\MSbar/Z_V$ and remove the discretization error.
\begin{eqnarray}\label{eq:fitZp_mom}
\frac{Z_P^\MSbar}{Z_V}(a^2p^2)&=&\frac{Z_P^\MSbar}{Z_V}+\sum_{i=1}^3C^{P,M}_i(a^2p^2)^i.
\end{eqnarray}
We take the fit region to be $p^2\in[9\,\GeV^2,18/a^2]$ and the corresponding fit result is 0.9675(20).  Comparing the coefficients in Table~\ref{tab:coff_zsmom} and Table~\ref{tab:coff_zpmom}, one can see that the discretization error in the pseudoscalar case is a bit larger than in the scalar case. The estimation of the systematic errors of $Z_P^\MSbar/Z_V$ is similar to the $Z_S$ case, while the nonzero strange quark mass effect is much larger, as shown in Table~\ref{tab:sea_dep_ZS}. The final result of $Z_P^\MSbar/Z_V$ at 2\,GeV is 0.968(2)(21)(16) and agrees with $Z_S^\MSbar/Z_V$ well; the three uncertainties here also correspond to the statistical error, systematic errors from the conversion ratio and other sources, respectively.

Now we turn to the SMOM case, which is free of the unphysical pole and has a trivial chiral extrapolation. The results of $Z_S^\MSbar/Z_V$ can also be obtained by using the matching factor between the $\SMOM$ and $\MSbar$ schemes up to three loops~\cite{Almeida:2010ns,Gorbahn:2010bf,Kniehl:2020sgo},
\begin{eqnarray}
&&C_S^{\MSbar,\SMOM}=\frac{Z_S^\MSbar/Z_V^\MSbar}{Z_S^\SMOM/Z_V^\SMOM}=\frac{Z_m^\SMOM}{Z_m^\MSbar}= \non\\
&&1+0.6455\left(\frac{\alpha_s}{4\pi}\right) +(23.0244-4.0135n_f)\left(\frac{\alpha_s}{4\pi}\right)^2+  \non\\
&&(889.736-169.924n_f+2.1844n_f^3)\left(\frac{\alpha_s}{4\pi}\right)^3
+\mathcal{O}(\alpha_s^4).
\end{eqnarray}
After converting the result of $\SMOM$ into the $\MSbar$ scheme and perturbatively running it to 2\,$\GeV$, we obtain the results in the right panel of Fig.~\ref{fig:Zs_momentum}. Compared to the results from the $\MOM$ scheme,
the results calculated with the $\SMOM$ scheme have better convergence in the perturbative matching. However, they also have stronger nonlinear dependence on $a^2p^2$ than that obtained through the $\MOM$ scheme. In order to describe the lattice data, we use the following two different models \cite{Liang:2021pql} to fit the data,
\begin{eqnarray}\label{eq:ZS_extrapo1}
\frac{Z_S^\MSbar}{Z_V}(a^2p^2)&=&\frac{Z_S^\MSbar}{Z_V}+\sum_{i=1}^3C_i^{S,S}(a^2p^2)^i, 
\end{eqnarray}
and also
\begin{eqnarray}\label{eq:ZS_extrapo2}
\frac{Z_S^\MSbar}{Z_V}(a^2p^2)&=&\frac{Z_S^\MSbar}{Z_V}+\frac{C_{-1}^{S,S}}{a^2p^2}+\sum_{i=1}^3C_i^{S,S}(a^2p^2)^i.
\end{eqnarray}
The fit results of $Z_S^\MSbar/Z_V$ and $\chi^2$/d.o.f.\ for different fit models and fit regions are listed in Table~\ref{tab:ZS_result_S}. Note that the $1/(a^2p^2)$ term is not applied to the fit in the $\MOM$ case since the results from the $\MOM$ scheme are also linearly dependent on $a^2p^2$ in the smaller $a^2p^2$ region with decreasing lattice spacing. One can anticipate that the nonperturbative physics region is related to $p^2$ rather than $1/(a^2p^2)$.
Compared to the fit without the $1/(a^2p^2)$ term, the fit with such a term can describe the data with much smaller $a^2p^2$ when we require $\chi^2$/d.o.f.\,$<$1.1, but the central value can be quite different. Since the $1/(a^2p^2)$ term reflects the nonperturbative effect in $Z_S$ with unknown origin, we have chosen the result fitted by Eq.~(\ref{eq:ZS_extrapo1}) with range $a^2p^2\in [3.5,9]$ to be the central value. Then we use the result with {the ansatz Eq.~(\ref{eq:ZS_extrapo1}) in the range} $a^2p^2\in [2.5,8]$ to estimate the systematic error caused by the fit range, and take the deviation between the central value and the result fitted by Eq.~(\ref{eq:ZS_extrapo2}) in the range $a^2p^2\in[1,9]$ as a systematic error. In summary, the $Z_S^\MSbar/Z_V$ at 2\,GeV through the RI/SMOM scheme is 0.964(6)(2)(51), with the latter two uncertainties from the conversion ratio and the other systematic uncertainties. With $n_f$=3, the matching coefficient $C_S^\SMOM$ can be written as
\begin{eqnarray}
C_{S, n_f=3}^{\MSbar,\SMOM}&=&1+0.0514\alpha_s+0.0696\alpha_s^2+0.2014\alpha_s^3  \nonumber\\
&&+\mathcal{O}(\alpha_s^4), 
\end{eqnarray}
and the truncation error from the matching is much smaller than in the RI/MOM case. 
 
\begin{table}[htbp]
  \centering
  \begin{tabular}{c|c|cc}
  \toprule
Fit ansatz & $a^2p^2$ range & Result & $\chi^2$/d.o.f.  \\
\hline
\multirow{3}{*}{Eq.(\ref{eq:ZS_extrapo1})} 
 &[2.5,8.0]  & 0.9767(30) & 0.2  \\
 &[3.5,9.0]  & 0.9643(57) & 0.3   \\
 &[3.5,10.5] & 0.9538(34) & 0.8  \\
\hline
\multirow{2}{*}{Eq.(\ref{eq:ZS_extrapo2})} 
&[1.0,8.0] & 0.9208(70)  & 0.7\\
&[1.0,9.0] & 0.9151(57)  & 0.8\\
\hline
\hline
  \end{tabular}
\caption{The fit results of $Z_S^\MSbar/Z_V$(2\,$\GeV$) {through the RI/SMOM scheme on the 64I ensemble} for different fit ranges of $a^2p^2$.}
  \label{tab:ZS_result_S}
\end{table} 

\begin{table}[htbp]
  \centering
  \begin{tabular}{c|c|cc}
  \toprule
Fit ansatz & Fit Range for $a^2p^2$ & Result & $\chi^2$/d.o.f. \\
\hline
\multirow{3}{*}{Eq.(\ref{eq:ZS_extrapo1})} 
 &[3.0,8.0]  & 0.9803(48) & 1.1 \\
 &[3.5,9.0]  & 0.9631(61) & 1.4\\
 &[3.5,10.5] & 0.9490(38) & 2.1\\
\hline
\hline
\multirow{2}{*}{Eq.(\ref{eq:ZS_extrapo2})} 
 &[1.0,9.0] & 0.9493(58)  & 2.2\\
 &[1.8,9.0] & 0.9130(120)  & 1.5 \\
\hline
\hline
  \end{tabular}
\caption{The fit results of $Z_P^\MSbar/Z_V$(2\,$\GeV$) {through the RI/SMOM scheme on the 64I ensemble}, for different fit ranges of $a^2p^2$.}
  \label{tab:ZP_result_S}
\end{table} 

The results of $Z_P^\MSbar/Z_V(2\,\GeV)$ from the $\SMOM$ scheme are presented in the right panel of Fig.~\ref{fig:Zp_momentum}. Even though $Z_P$ and $Z_S$ are very close to each other under the SMOM scheme at large $a^2p^2$, their difference at small $a^2p^2$ makes the acceptable fit range with reasonable $\chi^2$/d.o.f.\ to be different, as shown in Table~\ref{tab:ZP_result_S}. With similar analysis, we determine $Z_P^\MSbar/Z_V$ at 2\,GeV through the RI/SMOM scheme to be 0.963(6)(2)(53), which is consistent with $Z_S^\MSbar/Z_V$ within the uncertainty and the largest uncertainty comes from the fit model.

\subsection{Tensor current renormalization}

\begin{figure}[]
\begin{center}
	\includegraphics[scale=0.65]{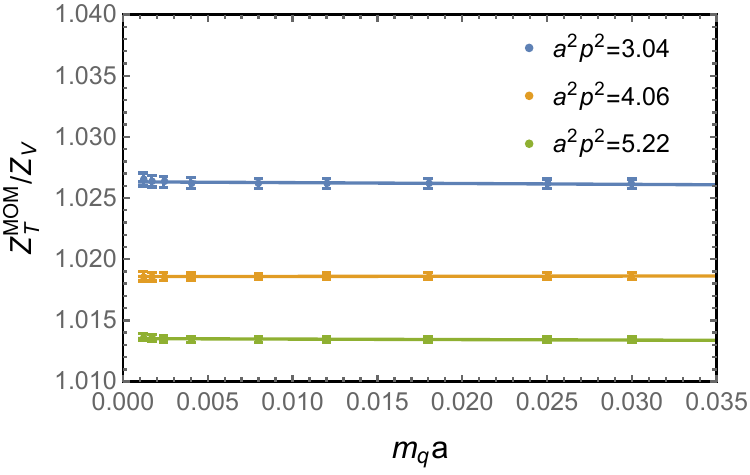}
	\includegraphics[scale=0.65]{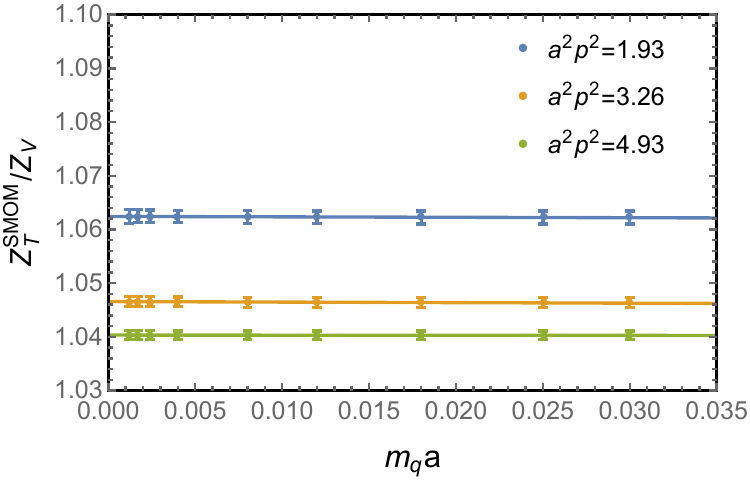}
\caption{The valence quark mass dependence of $Z^\MOM_T/Z_V$ (top panel) and of $Z^\SMOM_T/Z_V$ (bottom panel) at different $a^2p^2$ on the 64I ensemble. }
\label{fig:mass_Zt}
\end{center}
\end{figure}

The ratios of the RC of the tensor operator to $Z_V$ in the $\MOM$ and $\SMOM$ schemes can be obtained by 
\begin{eqnarray}
\frac{Z_T^\MOM}{Z_V^\MOM}=\frac{\Gamma_V(p)}{\Gamma_T(p)}\bigg|_{p^2=\mu^2}, \non\\
\frac{Z_T^\SMOM}{Z_V^\SMOM}=\frac{\Gamma_V(p_1,p_2)}{\Gamma_T(p_1,p_2)}\bigg|_{sym},
\end{eqnarray}
respectively, where
\begin{eqnarray}
&&\Gamma_T(p)=\frac{1}{144}\Tr[\Lambda^{\mu\nu}_{T,B}(p)\sigma_{\mu\nu}], \non\\
&&\Gamma_T(p_1,p_2)=\frac{1}{144}\Tr[\Lambda^{\mu\nu}_{T,B}(p_1,p_2)\sigma_{\mu\nu}].
\end{eqnarray}
The valence quark dependence of $Z_T^\MOM/Z_V$ and $Z_T^\SMOM/Z_V$
are plotted in Fig.~\ref{fig:mass_Zt}. 

\begin{figure}[]
 \begin{center}
	\includegraphics[scale=0.65]{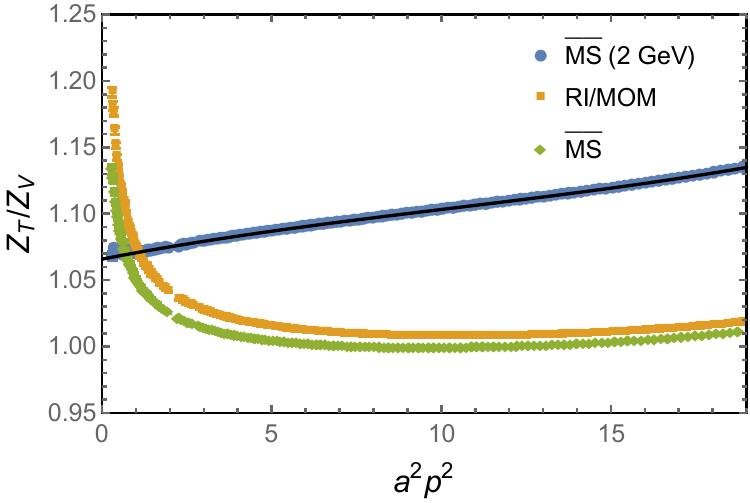}
	\includegraphics[scale=0.65]{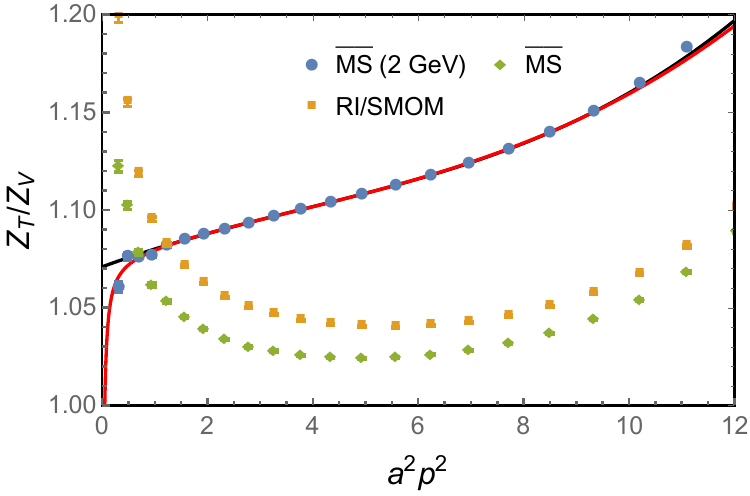}
\caption{Conversion and running of $Z_T^\MSbar/Z_V$ from the $\MOM$ scheme (top panel) and $\SMOM$ scheme (lower panel) on the 64I ensemble. The yellow data represent the results in the $\MOM$ (top panel) and the $\SMOM$ schemes (lower panel).  The green data are the results in the $\MSbar$ scheme with  $\mu=|p|$. The results in the $\MSbar$ scheme after running to 2 GeV are shown by the blue data. The black curves represent the fits using Eq.~(\ref{eq:fitZt_mom}) with data in $p^2\in[9\,\GeV^2,18/a^2]$ and Eq.~(\ref{eq:Zt_extrapo1}) with data in $a^2p^2\in[1.5,9]$ for the $\MOM$ and $\SMOM$ schemes, respectively. The red curve in the lower panel represents the fit using Eq.~(\ref{eq:Zt_extrapo2}) with data in $a^2p^2\in[0.3,9]$.}
\label{fig:Zt_momentum}
\end{center}
\end{figure}

As is the case for the quark field renormalization constant, the dependence of $Z_T/Z_V$ on the valence quark mass is mild {under both the RI/MOM and RI/SMOM schemes}, so we use linear extrapolation to obtain the $\MOM$ results in the chiral limit. {The solid lines in Fig.~\ref{fig:mass_Zt} are the fits with linear extrapolation and agree with the data well.} $Z_T^\MOM/Z_V$ and $Z_T^\SMOM/Z_V$ can be converted into the $\MSbar$ scheme with the matching factors~\cite{Almeida:2010ns,Gracey:2011fb,Kniehl:2020sgo},
\begin{eqnarray}\label{eq:zt_match_mom}
&&C_T^{\MSbar,\MOM}=1-\frac{1}{81}(4866-1656\zeta_3-259n_f)\left(\frac{\alpha_s}{4\pi}\right)^2
\non\\
&&+\frac{1}{17496}(21770010\zeta_3+231552\zeta_4-6505920\zeta_5- 46437951
\non\\
&&-1218240n_f\zeta_3+155520n_f z_4 + 5421360 n_f-6912n_f^2 \zeta_3
\non\\
&&-76336n_f^2)\left(\frac{\alpha_s}{4\pi}\right)^3+\mathcal{O}(\alpha_s^4),  
\end{eqnarray}
for the $\MOM$ scheme and
\begin{eqnarray}\label{eq:zt_match_smom}
&&C_T^{\MSbar,\SMOM}=1-0.21517295\left(\frac{\alpha_s}{4\pi}\right)-(43.38395
\non\\
&&-4.103279n_f)\left(\frac{\alpha_s}{4\pi}\right)^2+(-1950.76129+309.82858n_f
\non\\
&&-7.06359n_f^2)\left(\frac{\alpha_s}{4\pi}\right)^3
+\mathcal{O}(\alpha_s^4),
\end{eqnarray}
for the $\SMOM$ scheme.
The anomalous dimension of the tensor operator in the $\MSbar$ scheme, $\gamma_T^\MSbar$, has been calculated up to four loops in Landau gauge~\cite{Baikov:2006ai}. Then we obtain the results of $Z_T^\MSbar/Z_V$ at 2\,GeV from the intermediate schemes. Similar to other cases, we plot in Fig.~\ref{fig:Zt_momentum} the results in the $\MOM$ scheme (yellow data), in the $\MSbar$ scheme at $\mu=p$ (green data) and at 2\,GeV (blue data).
We use the following expression to fit the lattice data from the intermediate $\MOM$ scheme,
\begin{equation}\label{eq:fitZt_mom}
\frac{Z_T^\MSbar}{Z_V}(a^2p^2)=\frac{Z_T^\MSbar}{Z_V}+\sum_{i=1}^3C^{T,M}_i(a^2p^2)^i,
\end{equation}
and the following expressions
\begin{subequations}
\begin{eqnarray}
\label{eq:Zt_extrapo1}
\frac{Z_T^\MSbar}{Z_V}(a^2p^2)&=&\frac{Z_T^\MSbar}{Z_V}+\sum_{i=1}^3C^{T,S}_i(a^2p^2)^i, \\
\label{eq:Zt_extrapo2}
\frac{Z_T^\MSbar}{Z_V}(a^2p^2)&=&\frac{Z_T^\MSbar}{Z_V}+\frac{C_{-1}^{T,S}}{a^2p^2}+\sum_{i=1}^3C^{T,S}_i(a^2p^2)^i, \non\\
\end{eqnarray}
\end{subequations}
to fit the results from the $\SMOM$ scheme, which is similar to what we did in the analysis of other renormalization constants. Such a pole effect also has been observed in another calculation~\cite{Hasan:2019noy}. 
The fit results with different fit ansatzes and fit regions are shown in Table~\ref{tab:zt_ap_dep}. 

\begin{table}[htbp]
  \centering
  \begin{tabular}{c|cccc}
  \toprule
Ensemble & $C_1^{T,M}$ & $C_2^{T,M}$ &  $C_3^{T,M}$ \\
\hline
HISQ12L & 0.00555(13) & $-0.000192(12)$ & $6.28(34)\times10^{-6}$ \\
\hline
HISQ12H & 0.00475(26) & $-0.000130(27)$ & $4.91(91)\times10^{-6}$ \\
\hline
HISQ09 & 0.00541(06) & $-0.000199(06)$ & $6.09(20)\times10^{-6}$ \\
\hline
HISQ06 & 0.00554(13) & $-0.000242(14)$ & $7.10(51)\times10^{-6}$ \\
\hline
HISQ04 & 0.00574(04) & $-0.000269(04)$ & $7.45(14)\times10^{-6}$ \\
\hline
48I & 0.00462(12) & $-0.000142(11)$ & $5.42(34)\times10^{-6}$ \\
\hline
64I & 0.00496(05) & $-0.000180(05)$   & $5.80(16)\times10^{-6}$ \\
\hline
48If & 0.00526(03) & $-0.000209(04)$ & $6.27(11)\times10^{-6}$ \\
\hline
32If & 0.00595(06) & $-0.000272(07)$  & $7.78(24)\times10^{-6}$ \\
\hline
\hline
  \end{tabular}
\caption{The fit results of coefficients for the $\MOM$ scheme; the corresponding fit ansatz is Eq.~(\ref{eq:fitZt_mom}).} 
  \label{tab:coff_ztmom}
\end{table}

In the RI/MOM case, the fit range is chosen to be $p^2\in[9\,\GeV^2,18/a^2]$ and the result of $Z_T^\MSbar/Z_V$ is more insensitive to the range selection than other renormalization constants since the absolute values of the fit results for $C^{T,M}_i$ listed in Table~\ref{tab:coff_ztmom} are smaller than those of other operators; however, this is not the case for the $Z_T^\MSbar/Z_V$ through the RI/SMOM scheme due to much larger $a^2p^2$ dependence, as shown in Table~\ref{tab:zt_ap_dep}.

\begin{table}[htbp]
  \centering
  \begin{tabular}{c|c|cc}
  \toprule
Fit ansatz & Fit Range for $a^2p^2$ & Result & $\chi^2$/.  \\
\hline
\multirow{3}{*}{Eq.(\ref{eq:Zt_extrapo1})} 
&[1.0,8.0] & 1.0714(18)  & 0.09 \\
&[1.5,9.0] & 1.0709(25) & 0.09\\
&[3.5,10.5] & 1.0670(74) & 0.02\\
\hline
\hline
\multirow{1}{*}{Eq.(\ref{eq:Zt_extrapo2})} 
&[0.3,9.0] & 1.0758(26)  & 0.85 \\
\hline
\hline
  \end{tabular}
\caption{The fit results of $Z_T^\MSbar/Z_V$(2\,$\GeV$) through the $\SMOM$ scheme on the 64I ensemble.}
  \label{tab:zt_ap_dep}
\end{table}

When $n_f=3$, the conversion functions in Eq.~(\ref{eq:zt_match_mom}) and Eq.~(\ref{eq:zt_match_smom}) can be rewritten as
\begin{eqnarray}
C_{T, n_f=3}^\MOM&=&1-0.1641\alpha_s^2-0.4364\alpha_s^3+\mathcal{O}(\alpha_s^4), \\
C_{T, n_f=3}^\SMOM&=&1-0.0171\alpha_s-0.1968\alpha_s^2-0.5467\alpha_s^3\nonumber\\
&&+\mathcal{O}(\alpha_s^4).
\end{eqnarray}
Thus the estimate of the $\mathcal{O}(\alpha_s^4)$ coefficient $0.4364^2/0.1641=1.1605$ in the RI/MOM case is smaller than that in the SMOM case, $0.5467^2/0.1968=1.5187$. 

{Finally we get $Z_T^\MSbar/Z_V$ to be 1.0658(1)(43)(9) through the RI/MOM scheme and 1.071(2)(9)(6) through the RI/SMOM scheme. The uncertainty in the first bracket is the statistic error; the latter two uncertainties are from the conversion ratio and the other systematic uncertainties. The truncation error in the RI/SMOM case is smaller compared with our previous estimation in~\cite{Bi:2017ybi} by using the 3-loop result from Ref.~\cite{Kniehl:2020sgo}, but the sensitivity to the fit range is still much larger compared to the RI/MOM case. The nonzero strange quark mass effect is also estimated using the 24I ensembles and shown in Table~\ref{tab:sea_dep_ZT}. It turns out to be smaller than those of the other RCs.}

\begin{table}[htbp]
  \centering
  \begin{tabular}{c|ccccc}
  \toprule
$m_la$ & 0.005 & 0.01 & 0.02 & 0.03  \\
\hline
$Z_T^\MSbar/Z_V(2\,\GeV)$ & 1.0504(1)& 1.0505(1)& 1.0509(1) & 1.0512(1) \\
\hline
$\tilde{Z}_T^\MSbar/Z_V(4\,\GeV)$ & 1.0919(1)& 1.0921(1)& 1.0929(1) & 1.0935(1) \\
\hline
\hline
  \end{tabular}
\caption{The results of $Z_T^\MSbar/Z_V(2\,\GeV)$ and $Z_T^\MSbar/Z_V(4\,\GeV)$ on the 24I ensemble from the intermediate $\MOM$ and $\SMOM$ schemes, respectively. The corresponding slopes from the linear extrapolation of the light sea quark are 0.012(1) and 0.023(1) $\GeV^{-1}$. }
  \label{tab:sea_dep_ZT}
\end{table}

\section{Results}

\begin{table*}[htbp]
  \centering
  \begin{tabular}{l|cc|cc|cc|cc}
  \toprule
 & \multicolumn{2}{|c}{$Z_q^\MSbar(2\,\GeV)$} &\multicolumn{2}{|c}{$Z_S^\MSbar(2\,\GeV)$} & \multicolumn{2}{|c}{$Z_P^\MSbar(2\,\GeV)$} &\multicolumn{2}{|c}{$Z_T^\MSbar(2\,\GeV)$}   \\
 Ensemble & MOM & SMOM & MOM & SMOM & MOM & SMOM & MOM & SMOM \\
\hline
HISQ12L &1.233(06)  & 1.214(28)  & 1.221(06)(39)(15) & 1.181(65) &  1.274(13)(39)(40) & 1.236(75) & 1.150(07) & 1.147(58) \\
\hline
HISQ12H &1.245(07)  & 1.227(31) & 1.230(11)(40)(21) & 1.258(87) &  1.273(33)(40)(53) & 1.172(43) & 1.155(07) &  1.156(48)\\
\hline
HISQ09H &1.190(05) & 1.175(12)& 1.057(02)(25)(07) &  1.050(59) &  1.075(04)(24)(18) &  1.055(46) &  1.149(05) & 1.153(13)  \\
\hline
HISQ06 &1.152(04) & 1.148(12) & 0.946(02)(15)(05) & 0.962(21) &  0.950(04)(14)(18) & 0.961(22) & 1.154(04) & 1.160(08)   \\
\hline
HISQ04 &1.130(03) & 1.125(09) & 0.894(01)(10)(06) & 0.893(21) &  0.898(01)(09)(16) & 0.892(24) & 1.160(04) & 1.162(05)  \\
\hline
24D    & 1.364(24) &--& 1.407(07)(51)(14) &--& 1.426(16)(51)(30) &--& 1.229(9)   &- \\
\hline
24DH    & 1.368(23) &--& 1.426(08)(51)(21) &--& 1.452(24)(51)(49) &--& 1.235(9)  &-  \\
\hline
32Dfine    & 1.298(11) &--& 1.212(06)(54)(19) &--& 1.248(16)(54)(26) &--& 1.180(08)  &-  \\
\hline
48I    & 1.233(06) & 1.220(27)& 1.133(02)(37)(09) & 1.151(57) & 1.152(06)(36)(20) & 1.153(61) & 1.156(07) & 1.163(30)    \\
\hline
64I    & 1.188(05) & 1.175(16)& 1.034(01)(24)(06) & 1.040(55) & 1.044(02)(22)(17) & 1.039(58) & 1.150(05)  & 1.155(12)\\
\hline
48If   & 1.166(04) & 1.159(12)& 0.991(01)(19)(05) & 1.001(45) & 1.008(02)(18)(18) & 0.985(59) & 1.150(04) & 1.156(10) \\
\hline
32If   & 1.149(04) & 1.140(12)& 0.965(01)(18)(05) & 0.970(50) & 0.974(02)(17)(16) & 0.971(40) & 1.149(04) & 1.152(09)\\
\hline
  \end{tabular}
  \caption{The RCs of the scalar, pseudoscalar and tensor operators in the $\MSbar$ scheme at $\mu$=2~GeV obtained through the $\MOM$ and $\SMOM$ schemes. For the convenience of the quark mass and matrix element calculations, we separate the uncertainties of $Z_{S/P}$ through the $\MOM$ scheme into three parts: statistical error, systematic error from the fixed order truncation in the conversion ratio, and the other systematic errors listed in Table~\ref{tab:sum_error}. The first and third errors are independent across different ensembles, while the second one is correlated and will be suppressed in the continuum extrapolation.}
  \label{tab:result_mom}
\end{table*}

\begin{table*}[htbp]
  \centering
\setlength{\tabcolsep}{0.5mm}{
  \begin{tabular}{l|ccc|ccc|ccc|ccc|}
  \toprule
 & \multicolumn{3}{c|}{$Z_q^\MSbar(2\,\GeV)/Z_V$} & \multicolumn{3}{c|}{$Z_S^\MSbar(2\,\GeV)/Z_V$} & \multicolumn{3}{c|}{$Z_P^\MSbar(2\,\GeV)/Z_V$} &\multicolumn{3}{c|}{$Z_T^\MSbar(2\,\GeV)/Z_V$}   \\
\hline 
Ensemble & Range & Results & $\chi^2$/d.o.f.  &Range& Results & $\chi^2$/d.o.f.  & Range& Results & $\chi^2$/d.o.f.  & Range & Results & $\chi^2$/d.o.f.    \\
\hline 
HISQ12L & [3.4,18] & 1.1117(13)  & 0.21 & [3.4,18] & 1.101(05) & 0.21 & [3.4,18] & 1.149(12) & 0.09 & [3.4,18] & 1.0375(04)  & 1.10   \\
HISQ12H & [3.4,18] & 1.1214(15) & 0.31 & [3.4,18] & 1.083(10) & 0.32 & [3.4,18] & 1.147(30) & 0.06 & [3.4,18] & 1.0413(09) & 0.83   \\
HISQ09H & [1.8,18] & 1.0986(05)  & 0.56 & [1.8,18] & 0.975(02) & 0.24 & [1.8,18] & 0.992(04) & 0.10 & [1.8,18] & 1.0604(02) & 1.50   \\
HISQ06 & [0.8,18] & 1.0847(11) & 0.09 & [0.8,18] & 0.891(02)  & 0.01 & [0.8,18] & 0.895(04) & 0.01 & [0.8,18] & 1.0864(03) &  0.11 \\
HISQ04 & [0.4,18] & 1.0742(24) & 1.30 & [0.4,18] & 0.850(01) & 0.10  & [0.4,18] & 0.853(01) & 0.05  & [0.4,18] & 1.1023(01) & 0.93  \\
24D & [7.0,10] & 1.1191(36) & 0.16 & [9.0,13] & 1.155(05) & 0.23 & [9.0,13]  & 1.170(13) & 0.05 & [9.0,13] & 1.0089(09) & 1.50  \\
24DH & [7.0,10] & 1.1178(60) & 0.04  & [9.0,13] & 1.165(07)  & 0.15 & [9.0,13] & 1.186(20) & 0.01 & [9.0,13] & 1.0090(13) & 0.58  \\
32Dfine & [5.0,18] & 1.1371(38) & 0.12 & [5.0,18] & 1.062(05) & 0.15 & [5.0,18] & 1.093(14) & 0.06 & [5.0,18] & 1.0339(11) & 1.40  \\
48I & [3.0,18] & 1.1177(11) & 0.18 & [3.0,18] & 1.026(02) & 0.16 & [3.0,18] & 1.044(05) & 0.04 & [3.0,18] & 1.0472(04)  & 0.45 \\
64I & [1.7,18] & 1.1009(05) &  0.68 & [1.7,18] & 0.959(01)  & 0.05 & [1.7,18] & 0.968(02)  & 0.05 & [1.7,18] & 1.0658(01)  &  0.42 \\
48If & [1.2,18] & 1.0899(03)  & 1.70 & [1.2,18] & 0.927(01) & 0.12 & [1.2,18] & 0.942(02) & 0.19 & [1.2,18] & 1.0745(01)  &  1.20 \\
32If & [0.9,18] & 1.0797(04) & 1.30 & [0.9,18] & 0.906(01) & 0.16 & [0.9,18] & 0.914(02) & 0.06 & [0.9,18] & 1.0793(02) &  1.20 \\
\hline
  \end{tabular}}
  \caption{The fit range, fit results and $\chi^2$/d.o.f.\ for the different RCs from the $\MOM$ scheme on the 12 gauge ensembles. For most of the ensembles here, the fit ranges are chosen to be $p^2\in [9\,\GeV^2,18/a^2]$ and the fit ansatzes are defined in Eq.~(\ref{eq:fitzq_mom}), Eq.~(\ref{eq:fitZs_mom}), Eq.~(\ref{eq:fitZp_mom}) and Eq.~(\ref{eq:fitZt_mom}) for the four renormalization constants, respectively. For the two largest lattice spacing ensembles (24D and 24DH), we apply linear extrapolations to remove the $a^2p^2$ dependence since the region $p^2\in [9\,\GeV^2,18/a^2]$ is not sufficiently large and the fit result will suffer large uncertainties if using the fit ansatz which includes the $(a^2p^2)^3$ term. The $\chi^2/\text{d.o.f.}$ of $Z_P^\MSbar(2\,\GeV)/Z_V$ are smaller than those of the other RCs since they have the largest statistical errors compared to the other RCs, which is caused by the contamination from the infrared physics mentioned in Sec.~\ref{sec:s_and_ps}.}
  \label{tab:sum_FR_mom}
\end{table*}

\begin{table*}[htbp]
  \centering
\setlength{\tabcolsep}{0.5mm}{
  \begin{tabular}{l|lcc|lcc|lcc|lcc|}
  \toprule
 & \multicolumn{3}{c|}{$Z_q^\MSbar(2\,\GeV)/Z_V$} & \multicolumn{3}{c|}{$Z_S^\MSbar(2\,\GeV)/Z_V$} & \multicolumn{3}{c|}{$Z_P^\MSbar(2\,\GeV)/Z_V$} &\multicolumn{3}{c|}{$Z_T^\MSbar(2\,\GeV)/Z_V$}   \\
\hline 
Ensemble & Range & Results & $\chi^2$/d.o.f.  & Range & Results & $\chi^2$/d.o.f.  & Range & Results & $\chi^2$/d.o.f.  & Range & Results & $\chi^2$/d.o.f.    \\
\hline 
HISQ12L & [1.0,9.0] & 1.0953(45) & 0.42 & [2.0,9.0] & 1.065(14)  & 0.36  & [2.0,9.0] & 1.115(15) & 0.85 & [1.0,9.0] & 1.0347 (43) & 1.30   \\
HISQ12H & [1.0,9.0] & 1.1064(37) & 0.04 & [2.0,9.0] & 1.134(23)  & 0.48  & [2.0,9.0] & 1.057(22) & 0.06 & [1.0,9.0] & 1.0426(28) & 0.12   \\
HISQ09 & [1.5,9.5] & 1.0845(36) & 0.07 & [3.5,9.5] & 0.969(05) & 0.07 & [3.5,9.5] & 0.974(05) & 0.44  & [1.5,9.5] & 1.0642(24) & 0.01   \\
HISQ06 & [1.0,9.0] & 1.0813(78) & 0.01 & [2.0,9.0] & 0.906(02)  & 1.00 & [2.0,9.0] & 0.905(02)  & 0.96 & [1.0,9.0] & 1.0923(41) & 0.01   \\
HISQ04 & [1.0,9.0] & 1.0690(23) & 0.04 & [2.5,9.0] & 0.848(02) & 0.97  & [2.5,9.0] & 0.847(02) & 0.38 & [1.0,9.0] & 1.1049(19)  & 0.01  \\
48I & [1.0,9.0] & 1.1056(37)  & 0.13 & [3.0,9.0] & 1.043(07) & 0.18 & [3.0,9.0] & 1.045(07) & 0.28 & [1.0,9.0] & 1.0536(27)  & 0.10  \\
64I & [1.8,9.0] & 1.0891(46) & 0.10 & [3.5,9.0] & 0.964(06) & 0.30  & [3.5,9.0] & 0.963(06)  & 1.40 & [1.0,9.0] & 1.0709(25)  & 0.09  \\
48If & [1.0,9.0] & 1.0836(22)  & 0.19  & [2.5,9.0] & 0.936(02) & 0.92 & [3.0,9.0] & 0.920(04) & 0.23 & [1.0,9.0] & 1.0803(17)  & 0.06  \\
32If & [1.5,9.5] & 1.0708(46) & 0.06 & [2.5,9.5] & 0.911(02) & 0.95 & [2.5,9.5] & 0.913(02)  & 1.80 & [1.5,9.5] & 1.0817(34)  & 0.02  \\
\hline
  \end{tabular}}
  \caption{The fit range, fit results and $\chi^2$/d.o.f.\ for the different RCs from the $\SMOM$ scheme on the 9 gauge ensembles.}
  \label{tab:sum_FR_smom}
\end{table*}

Following a similar strategy, the results of RCs on all the ensembles are listed in Table~\ref{tab:result_mom}, and the fit ranges used for the central values are collected in Tables~\ref{tab:sum_FR_mom} and \ref{tab:sum_FR_smom}. For the ensembles with lattice spacing smaller than 0.15~fm, we apply the same fit ansatz to extrapolate the results from the intermediate $\MOM$ scheme to the $a^2p^2\rightarrow0$ limit, and the corresponding fit region is $p^2\in[9\,\GeV^2,18/a^2]$. For the two largest lattice spacing ensembles (24D and 24DH), we apply linear extrapolation to remove the $a^2p^2$ dependence since the region $p^2\in [9\,\GeV^2,18/a^2]$ is not sufficiently wide and the fit results on these two ensembles will suffer large uncertainties if using the fit ansatz which includes the $(a^2p^2)^3$ term. The $\chi^2$/d.o.f.\ of the RCs in most of the cases are smaller than 1; for each much smaller than 1, one might have the concern of a possible overestimate of the statistical uncertainty of the RC. However, as shown in Tables~\ref{tab:sum_error} and~\ref{tab:sum_error_S}, the statistical uncertainty is much smaller than some of the systematic uncertainties. Therefore, such an overestimate will not change the total uncertainty. The SMOM scheme is not applied to the DSDR ensembles since the available data points are very limited due to the large lattice spacing and nonperturbative effects in the small $a^2p^2$ region. Generally speaking, both the statistical and systematic uncertainties are suppressed at smaller lattice spacing, since the calculation with higher momentum will have smaller quantum fluctuation and will thus be more precise. 

In $\MOM$ scheme, there is an unphysical mass pole in the calculation of each of $Z_S/Z_V$ and $Z_P/Z_V$. Thus, we have to use Eq.(\ref{eq:ZS_massfit}) and Eq.(\ref{eq:ZP_massfit}) to remove these mass poles and obtain the results in the chiral limit; however, for $Z_q/Z_V$ and $Z_T/Z_A$ we can do linear chiral extrapolations. The $\MSbar$ results of the quark bilinear operators through the $\MOM$ scheme are almost linearly dependent on $a^2p^2$. It is surprising that the deviation from the linear behavior  is still less than 1\% at $a^2p^2=14$, or more precisely, $a^2p^2_{\mu}\simeq a^2p^2/4=3.5$. It allows us to use the data at large $a^2p^2$ to suppress the influence of the nonperturbative effects, and guarantees a reliable $a^2p^2$ polynomial fit when we extract the renormalization constant through the $\MOM$ scheme. Ultimately, most of the RC uncertainties are due to the truncation error in the matching factors, which can be larger than 1\% for $Z_{S/P}$ for most of the lattice spacings. But such an uncertainty is correlated across all the lattice spacings, and is suppressed at smaller lattice spacing thanks to a larger fit range. Thus, we can separate the uncertainty of each RC into two pieces, that from the matching and the others, and treat them differently in the continuum extrapolation. For example, assuming the renormalized light quark masses at 0.114 and 0.084 fm lattice spacings are 3.34(4)(10) MeV and 3.34(4)(7) MeV respectively, with the second error from the matching (and fully correlated at the two lattice spacings) and the first error from the other sources independent at the two lattice spacings, 
then the final result will be something like 3.34(9)(5) MeV, where the first error is obtained by applying a linear fit to the results on these two ensembles, or alternatively through error propagation, and the second error is estimated assuming the truncation error in the matching of the RCs linearly decreases with decreasing lattice spacing.

In order to illustrate and compare the discretization errors of the RCs at different lattice spacings, we normalize $Z_{S,T}^{\MOM}(p^2;a)$ at different lattice spacings under the RI/MOM scheme with the corresponding $\overline{\mathrm{MS}}$ values $Z^\MSbar_{\mathcal{O}}(2\,\GeV;a)$ at 2\,GeV, with the following definitions:
\begin{align}
&Z^{\MOM, {\rm norm}}_\mathcal{O}(p^2;a)=Z^{\MOM}_\mathcal{O}(p^2;a)/Z^\MSbar_{\mathcal{O}}(2\,\GeV;a),\nonumber\\
&Z^{\MOM, {\rm sub}_n}_\mathcal{O}(p^2;a)=Z^{\MOM, {\rm norm}}_\mathcal{O}(p^2;a)\nonumber\\
&\quad \quad \quad -\frac{\sum_{i=1}^{n}C_i^{\mathcal{O},M}(a^2p^2)^i}{Z^\MSbar_{\mathcal{O}}(2\,\GeV;a) C^{\MOM}_{\mathcal{O}}(p^2)R^\MSbar_{\mathcal{O}}(|p|,2\,\GeV)},
\end{align}
where $R^\MSbar_{\mathcal{O}}(\mu_1,\mu_2)$ is the evolution ratio under the $\MSbar$ scheme from the scale $\mu_1$ to $\mu_2$. $Z^{\MOM, {\rm norm}}_\mathcal{O}$ can also be calculated under dimensional regularization and it is simply $Z^{\MOM, {\rm norm}}_\mathcal{O}(p^2;\epsilon)\equiv \big(C^\MOM_{\mathcal{O}}(p^2)R^\MSbar_{\mathcal{O}}(|p|,2\,\GeV)\big)^{-1}$, and $Z^{\MOM, {\rm sub}_n}$ describes the normalized RI/MOM renormalization constant when the discretization error up to $(a^2p^2)^n$ order is subtracted. 

\begin{figure*}[]
\begin{center}
\subfigure[]
{
	\begin{minipage}[b]{0.45\linewidth}
	\centering 
	\includegraphics[scale=0.6]{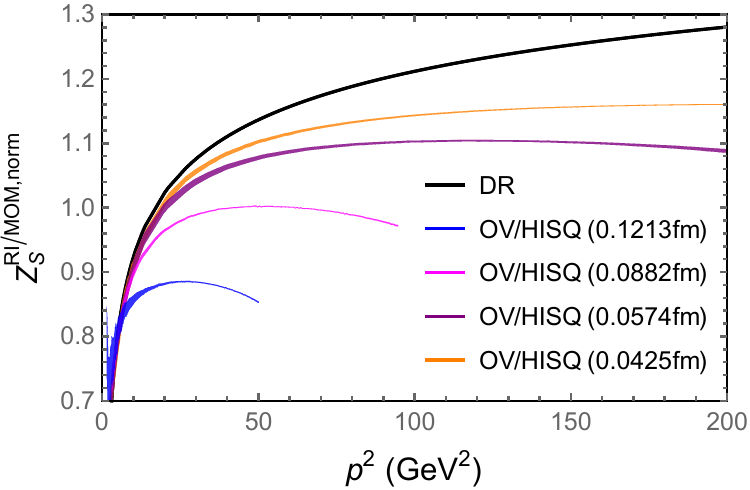}
	\end{minipage}
}
\subfigure[]
{
	\begin{minipage}[b]{0.45\linewidth}
	\centering  
	\includegraphics[scale=0.6]{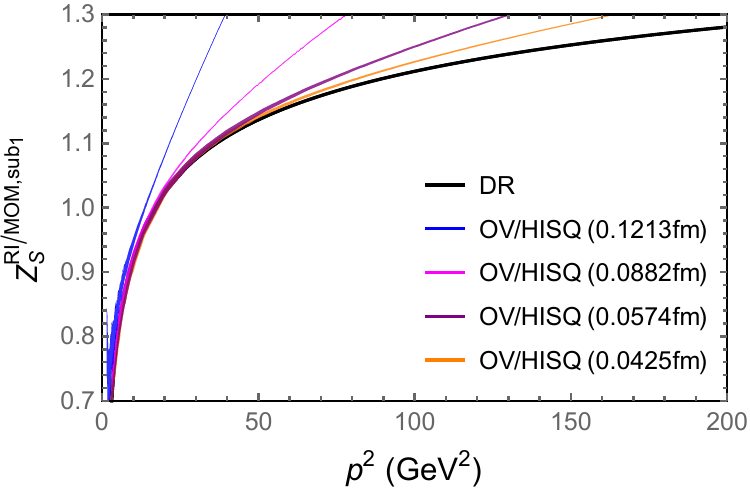}
	\end{minipage}
}
\subfigure[]
{
	\begin{minipage}[b]{0.45\linewidth}
	\centering 
	\includegraphics[scale=0.6]{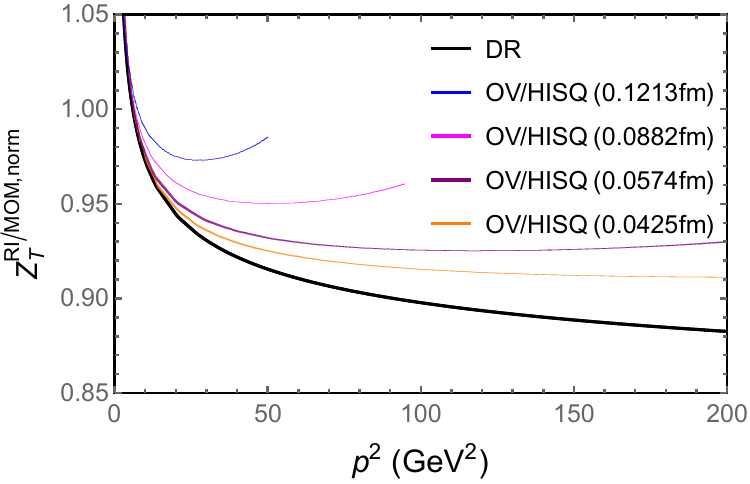}
	\end{minipage}
}
\subfigure[]
{
	\begin{minipage}[b]{0.45\linewidth}
	\centering  
	\includegraphics[scale=0.6]{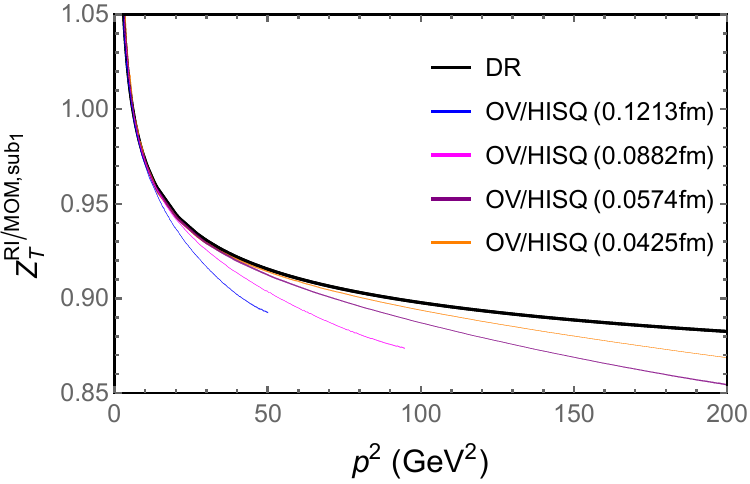}
	\end{minipage}
}
\caption{Normalized RI/MOM renormalization constants under the $\MSbar$ scheme (black lines) and lattice regularization (colored bands). The two left panels show the $Z_S$ (upper panel) and $Z_T$ (lower panel) cases based on the original lattice results, and the right panels show the cases with the  linear $a^2p^2$ correction subtracted.}
\label{fig:norm_mom}
\end{center}
\end{figure*}

Since the window we used for the $\MOM$ case covers all the data in the range $p^2>9\,\GeV^2$ and the discretization error is relatively small, we just compare the original $Z^{\MOM, {\rm norm}}_{S/T}(p^2;a)$ and the $Z^{\MOM, {\rm sub}_1}_{S/T}(p^2;a)$ with their counterpart $Z^{\MOM, {\rm norm}}_{S/T}(p^2;\epsilon)$ in the dimensional regularization, as shown in Fig.~\ref{fig:norm_mom}. One can see that the difference between $Z^{\MOM, {\rm norm}}_{S/T}(p^2;a)$ and $Z^{\MOM, {\rm norm}}_{S/T}(p^2;\epsilon)$ becomes smaller when the lattice spacing becomes smaller, and the subtraction of the linear $a^2p^2$ correction improves the convergence of the $Z^{\MOM, {\rm sub}_1}_{S/T}(p^2;a)$ significantly.

Compared with the $\MOM$ scheme cases, the effect of the unphysical mass pole is much smaller in the $\SMOM$ scheme. So we only choose the linear chiral extrapolation to obtain the results in the chiral limit. The perturbative convergence of the scalar and pseudoscalar operators in the $\SMOM$ scheme are better than that in the $\MOM$ scheme when matched to the $\MSbar$ scheme, while the situation is opposite for the tensor operator. After converting the results calculated with the $\SMOM$ scheme to $\MSbar$ 2\,GeV, the data show very strong dependence on $a^2p^2$; it leads to a large systematic error caused by the fit region of $a^2p^2$, and it contributes most of the uncertainty to $Z_q^\MSbar(2\,\GeV)$. For the renormalization constants of the quark field $Z_q^\MSbar(2\,\GeV)$ and tensor operator $Z_T^\MSbar(2\,\GeV)$, the effects of the  $1/(a^2p^2)$ pole are much smaller compared to those in the scalar and pseudoscalar cases.
For the results of the scalar and pseudoscalar operators, we find the form with an additional $1/(a^2p^2)$ term can have better description of the data at small $1/(a^2p^2)$, while the prediction will differ from that without this term. Most of the systematical errors of $Z_S^\MSbar(2~\GeV)$ and $Z_P^\MSbar(2~\GeV)$ come from this deviation. 

\begin{figure*}[]
\begin{center}
\subfigure[]
{
	\begin{minipage}[b]{0.45\linewidth}
	\centering    
	\includegraphics[scale=0.65]{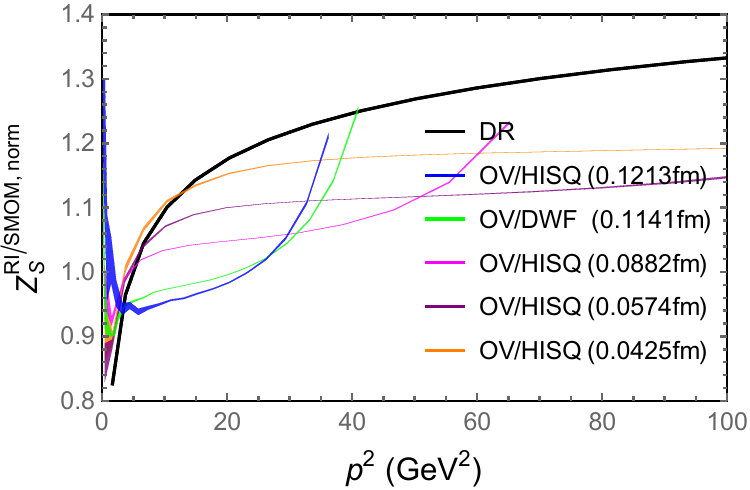} 
	\end{minipage}
}
\subfigure[] 
{
	\begin{minipage}[b]{0.45\linewidth}
	\centering      
	\includegraphics[scale=0.65]{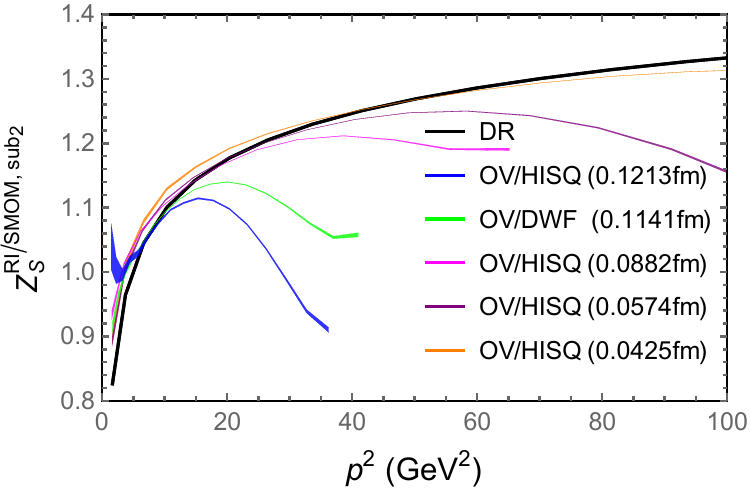} 
	\end{minipage}
}
\subfigure[]
{
	\begin{minipage}[b]{0.45\linewidth}
	\centering 
	\includegraphics[scale=0.65]{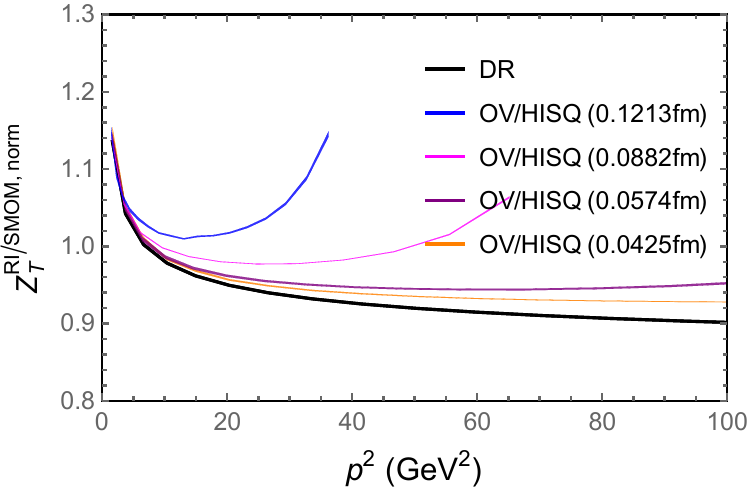}
	\end{minipage}
}
\subfigure[]
{
	\begin{minipage}[b]{0.45\linewidth}
	\centering  
	\includegraphics[scale=0.65]{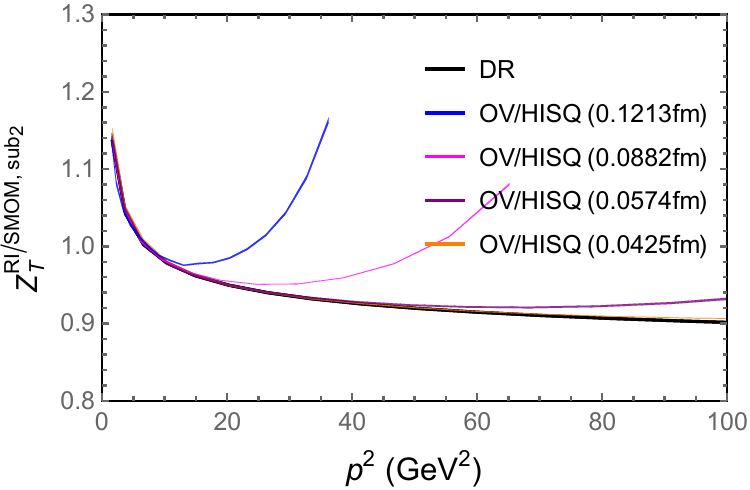}
	\end{minipage}
}
\caption{Normalized RI/SMOM renormalization constants under the $\MSbar$ scheme (black lines) and lattice regularization (colored bands). The curves in the Fig.~(a) show the large discrepancy between results under the lattice regularization and dimension regularization, especially the results on the relatively larger lattice spacing ensembles, which decrease and then increase with increasing $p^2$. The discretization errors shown in the right panels are larger than that in the RI/MOM case even after the linear and quadratic terms of $a^2p^2$ in $Z^{\SMOM}_\mathcal{O}(p^2;a)$ have been subtracted. From Fig.~(b), one can see that the differences between the dimensional regularized values $Z^{\SMOM, {\rm norm}}_S(p^2;\epsilon)$ (black bands) and the lattice regularized values $Z^{\SMOM, {\rm sub}_2}_S(p^2;a)$ (colored bands) at small $p^2$ become larger when the lattice spacing $a$ becomes smaller; the results on the smaller lattice spacing ensembles are consistent with DR result only when $p^2$ becomes larger, which can be seen by comparing the orange and black lines.}
\label{fig:norm_smom}
\end{center}
\end{figure*}

We can also make a similar comparison for the $\SMOM$ case for the discretization error, with the following definitions,
\begin{align}
&Z^{\SMOM, {\rm norm}}_\mathcal{O}(p^2;a)=Z^{\SMOM}_\mathcal{O}(p^2;a)/Z^\MSbar_{\mathcal{O}}(2\,\GeV;a),\nonumber\\
&Z^{\SMOM, {\rm sub}_n}_\mathcal{O}(p^2;a)=Z^{\SMOM, {\rm norm}}_\mathcal{O}(p^2;a)\nonumber\\
&\quad \quad \quad -\frac{\sum_{i=1}^{n}C_i^{\mathcal{O},S}(a^2p^2)^i}{Z^\MSbar_{\mathcal{O}}(2\,\GeV;a) C^\SMOM_{\mathcal{O}}(p^2)R^\MSbar_{\mathcal{O}}(|p|,2\,\GeV)}.
\end{align}
As shown in Fig.~\ref{fig:norm_smom}, the original lattice data of $Z^{\SMOM, {\rm norm}}$ have huge discretization errors, and the effect is still obvious even after the linear and quadratic terms of $a^2p^2$ are subtracted. At the same time, there is a sizable difference between $Z^{\SMOM, {\rm norm}}_S(p^2;\epsilon)\equiv \big(C^\SMOM_{\mathcal{O}}(p^2)R^\MSbar_{\mathcal{O}}(|p|,2\,\GeV)\big)^{-1}$ under dimensional regularization and $Z^{\SMOM, {\rm sub}_2}_\mathcal{O}(p^2;a)$ in the small $p^2$ region, which is illustrated in Fig.~\ref{fig:norm_smom}(b); one can see that the difference becomes larger approaching the continuum limit in the region $p^2\in[5\,\GeV^2,40\,\GeV^2]$. It suggests that there is an unknown effect which should be removed  before the accurate scalar renormalization constant can be extracted using the SMOM data in the small $p^2$ region.

\section{Summary}\label{sec:summary}

In this work we systematically studied the RCs of quark field $Z_q$ and bilinear quark operators ($Z_V$, $Z_S$, $Z_P$ and $Z_T$) using the intermediate $\MOM$ and $\SMOM$ schemes. We used the overlap valence quark on 2+1 DWF gauge ensembles and 2+1+1 HISQ gauge ensembles. 
The PCAC relation has been used to obtain the RC of axial vector current. The ratios of $Z_q$ to $Z_V$ were obtained by the bare amputated Green function of the axial vector operator.
The ratios of other RCs to $Z_V$ were obtained though the ratios of appropriate vertex functions. We converted the RCs to the $\MSbar$ scheme and used the corresponding anomalous dimensions to run the energy scale to 2\,GeV. After extrapolating the results to the $a^2p^2\rightarrow0$ limit, we obtained consistent results from the intermediate $\MOM$ and $\SMOM$ schemes. These results are summarized in Table~(\ref{tab:result_mom}). 

\begin{figure*}[]
\begin{center}
\subfigure[]
{
	\begin{minipage}[b]{0.45\linewidth}
	\centering 
	\includegraphics[scale=0.65]{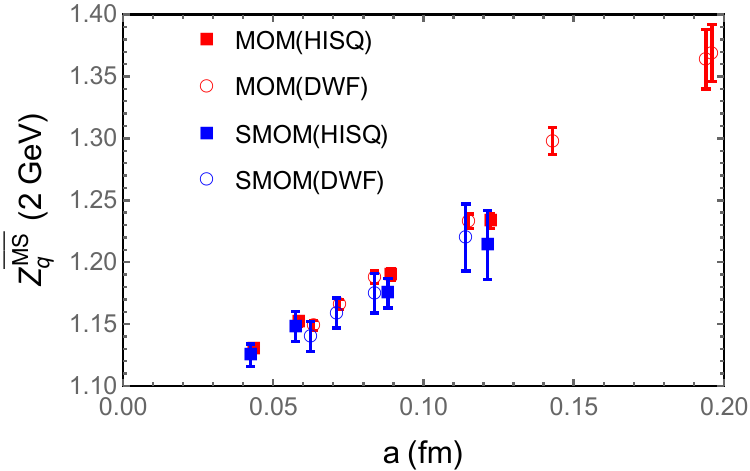}
	\end{minipage}
}
\subfigure[]
{
	\begin{minipage}[b]{0.45\linewidth}
	\centering  
	\includegraphics[scale=0.65]{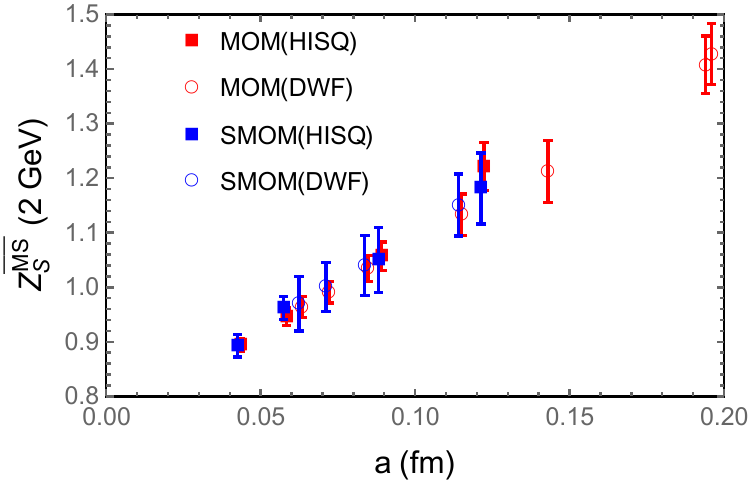}
	\end{minipage}
}
\subfigure[]
{
	\begin{minipage}[b]{0.45\linewidth}
	\centering    
	\includegraphics[scale=0.65]{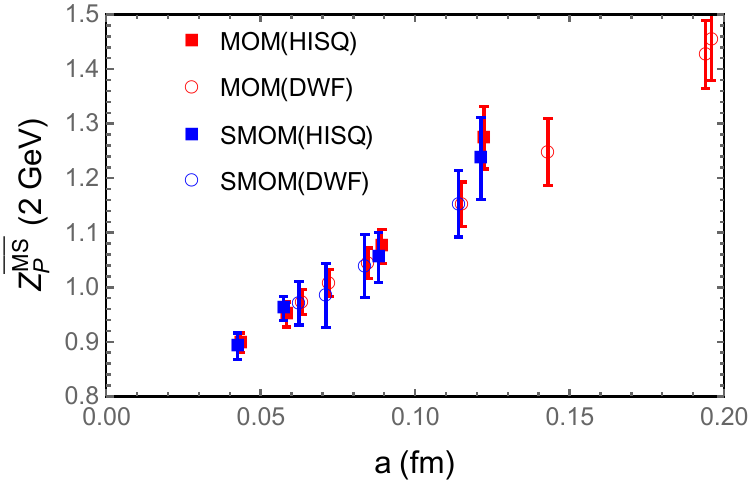}
	\end{minipage}
}
\subfigure[] 
{
	\begin{minipage}[b]{0.45\linewidth}
	\centering      
	\includegraphics[scale=0.65]{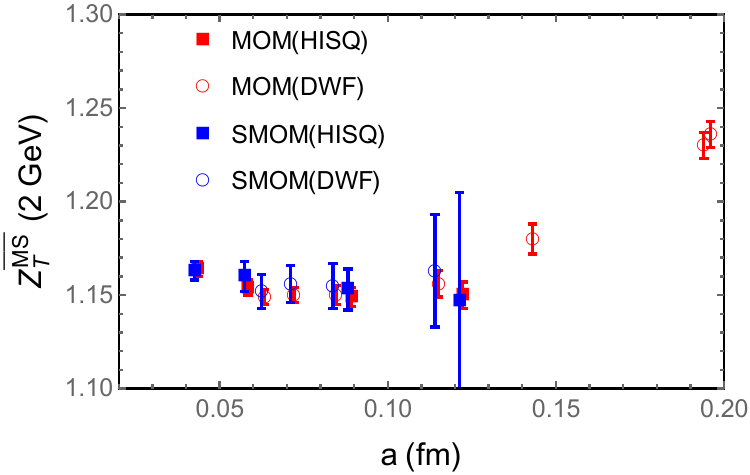}
	\end{minipage}
}
\caption{$Z_q^\MSbar(2\,\GeV)$, $Z_S^\MSbar(2\,\GeV)$, $Z_P^\MSbar(2\,\GeV)$ and $Z_T^\MSbar(2\,\GeV)$ on the different gauge ensembles. The red and blue data represent the RCs from the $\MOM$ and $\SMOM$ schemes, respectively. The filled boxes and open circles denote the the results on the HISQ ensembles and DWF ensembles, respectively.}
\label{fig:final_results}
\end{center}
\end{figure*}

We also present these results in Fig.~\ref{fig:final_results}. The red and blue data represent the RCs from the $\MOM$ and $\SMOM$ schemes, respectively; the filled boxes are the results on the HISQ ensembles while the open circles are the results on the DWF ensembles. Though the bare coupling constants are very different between the HISQ and DWF ensembles ($6/g^2\sim 3.6$ and 2.2 respectively at $a\simeq 0.1$ fm), our results show the RCs are more sensitive to the lattice spacing rather than the bare coupling constants. It means that the bare $g^2$ is not suitable to be used in the perturbative expansion. A more suitable coupling constant is very close to the one in the $\overline{\text{MS}}$ scheme~\cite{Lepage:1992xa}, which is sensitive to $\pi/a$ but not $g^2$. It also suggests that one can combine the renormalized overlap fermion results on both the HISQ and DWF ensembles to obtain a more reliable continuum limit. 

In the appendix, we show the preliminary results of the scalar and tensor renormalization constants using the interpolating momentum (IMOM) scheme~\cite{Sturm:2009kb,Gorbahn:2010bf,Garron:2021mfn} , with different momentum transfer factor $\omega\equiv (p_2-p_1)^2/p_1^2$. The results suggest that the results with different $\omega$ can be quite sensitive to $\omega$ even though they become closer at finer lattice spacing. Thus the schemes with non-zero momentum transfer can suffer from additional systematic uncertainties and require careful treatment, even though the perturbative convergence in certain cases is extremely good. We note that the 4-loop perturbative matching of the tensor and scalar operators from the $\RIp$ scheme to $\MSbar$ scheme has been obtained recently~\cite{Gracey:2022vqr,Gracey:2022vjc}; it shows the 4-loop correction for scalar operator is still large and will be very important to improve the precision of $Z_S$ from the RI/MOM scheme.


\section*{Acknowledgments}
We thank the RBC and UKQCD collaborations for providing us their DWF gauge configurations, the MILC collaboration for providing their HISQ gauge configurations, and J.A. Gracey for valuable discussions. The calculations were performed using the GWU code~\cite{Alexandru:2011ee,Alexandru:2011sc} through the HIP programming model~\cite{Bi:2020wpt}.  The numerical calculation is supported by the Strategic Priority Research Program of Chinese Academy of Sciences, Grant No.\  XDC01040100, and also the supercomputing system in the Southern Nuclear Science Computing Center (SNSC). 
This research used resources of the Oak Ridge Leadership Computing Facility at the Oak Ridge National Laboratory, which is supported by the Office of Science of the U.S. Department of Energy under Contract No.\ DE-AC05-00OR22725. This work used Stampede time under the Extreme Science and Engineering Discovery Environment (XSEDE), which is supported by National Science Foundation Grant No. ACI-1053575.
We also thank the National Energy Research Scientific Computing Center (NERSC) for providing HPC resources that have contributed to the research results reported within this paper.
We acknowledge the facilities of the USQCD Collaboration used for this research in part, which are funded by the Office of Science of the U.S. Department of Energy.
Y.B. and Z.L. are supported in part by the National Natural Science Foundation of China (NNSFC) under Grant No.\ 12075253 (Y.B., Z.L.) and 11935017 (Z.L.). T.D. and K.L. are supported by the U.S. DOE Grant No.\ DE-SC0013065 (T.D., K.L.) and DOE Grant No.\ DE-AC05-06OR23177 (K.L.), which is within the framework of the TMD Topical Collaboration. Y.Y. is supported by the Strategic Priority Research Program of Chinese Academy of Sciences, Grant No.\ XDB34030303, XDPB15 and a NSFC-DFG joint grant under Grant Nos.\ 12061131006 and SCHA~458/22.

\bibliographystyle{apsrev4-1}
\bibliography{condensate.bib}

\section*{Results through the interpolating-momentum scheme}
In the appendix, we provide preliminary results to renormalize the scalar quark operator using the interpolating momentum (IMOM) scheme~\cite{Sturm:2009kb,Gorbahn:2010bf,Garron:2021mfn}. The momenta in the IMOM scheme are chosen to be 
\begin{eqnarray}\label{eq:mom_IMOM}
p_1^2=p_2^2=\mu^2,~~~~~(p_2-p_1)^2=\omega\mu^2{.}
\end{eqnarray}
The value of $\omega$ ranges from 0 to 4, and $\omega=0$ and $\omega=1$ correspond to the $\MOM$ and $\SMOM$ schemes, respectively. The renormalization conditions in an IMOM scheme are similar as those in the SMOM scheme in Eq.~(\ref{eq:rismom}) except the momentum is set by Eq.~(\ref{eq:mom_IMOM}). There are two choices of momentum which can satisfy the condition~(\ref{eq:rismom}), as shown in Table~\ref{tab:mom_IMOM}. Note that on certain lattices such as HISQ12H ($L^3\times T=24^3\times 64$), the momenta which can be used with Scenario B are very limited and make a reliable result inaccessible. 

\begin{table}[htbp]
  \centering
 \resizebox{\linewidth}{!}{\begin{tabular}{c|c|c|c}
  \toprule
   & $\omega$ & $p_1$ & $p_2$ \\
\hline
\multirow{4}{*}{Scenario A} 
 & $\omega=1$ & ($q$,~$q$,~0,~0) & ($q$,~0,~$q$,~0) \\
 & $\omega=2$ & ($q$,~$q$,~0,~0) & (0,~0,~$q$,~$q$)   \\
 & $\omega=3$ & ($q$,~$q$,~0,~0) & (0,$-q$,~$q$,~0)  \\
 & $\omega=4$ & ($q$,~$q$,~0,~0) & ($-q$,$-q$,~0,~0) \\
 \hline
\multirow{4}{*}{Scenario B} 
 & $\omega=1$ & ($q$,~$q$,~$q$,~$q$) & ($-q$,~$q$,~$q$,~$q$) \\
 & $\omega=2$ & ($q$,~$q$,~$q$,~$q$) & ($-q$,$-q$,~$q$,~$q$)   \\
 & $\omega=3$ & ($q$,~$q$,~$q$,~$q$) & ($-q$,$-q$,$-q$,~$q$)  \\
 & $\omega=4$ & ($q$,~$q$,~$q$,~$q$) & ($-q$,$-q$,$-q$,$-q$) \\
 \hline
  \end{tabular}
}
\caption{Different momentum scenario of IMOM scheme with different $\omega$.}
  \label{tab:mom_IMOM}
\end{table} 

\begin{figure*}[]
\begin{center}
	\includegraphics[scale=0.65]{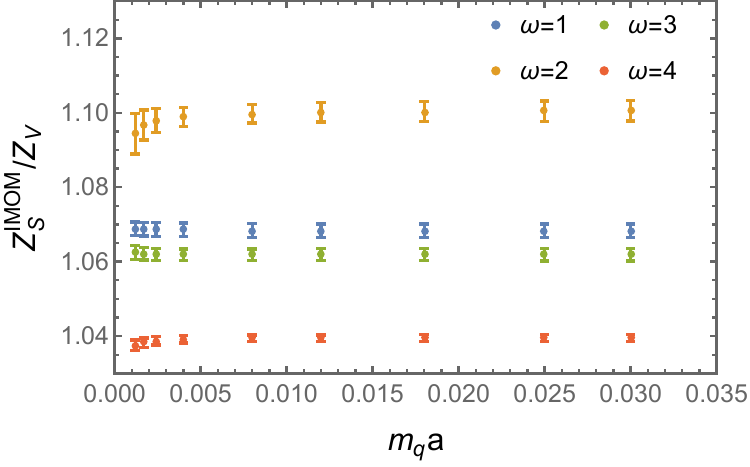}
	\includegraphics[scale=0.65]{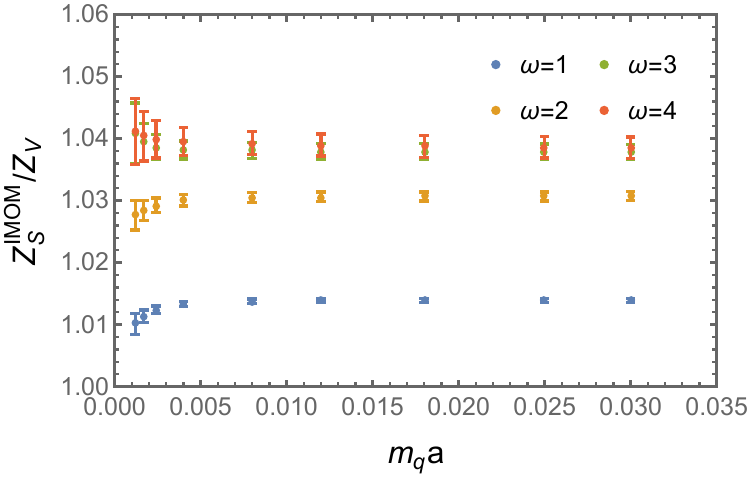}
\caption{The valence quark mass dependence of $Z_S^{\IMOM}/Z_V$ on the 64I ensemble with different $\omega$ when $a^2p^2=3.26$ for scenario A and $a^2p^2=3.13$ for scenario B.}
\label{fig:IMOM_mass_dep}
\end{center}
\end{figure*}

\begin{figure*}[]
\begin{center}
	\includegraphics[scale=0.65]{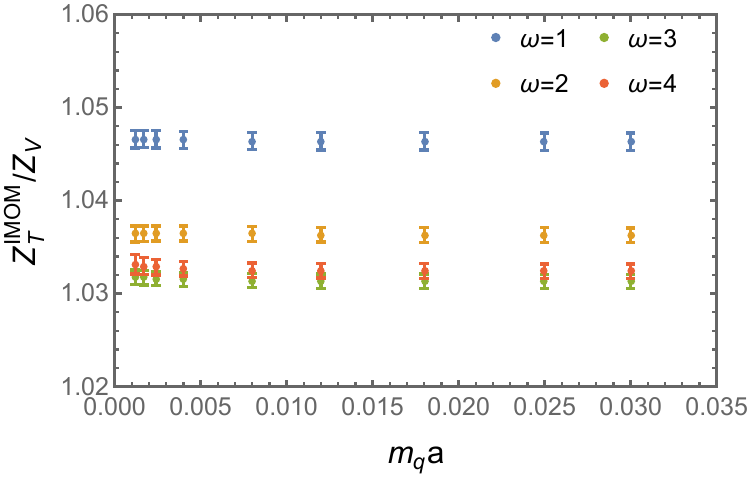}
	\includegraphics[scale=0.65]{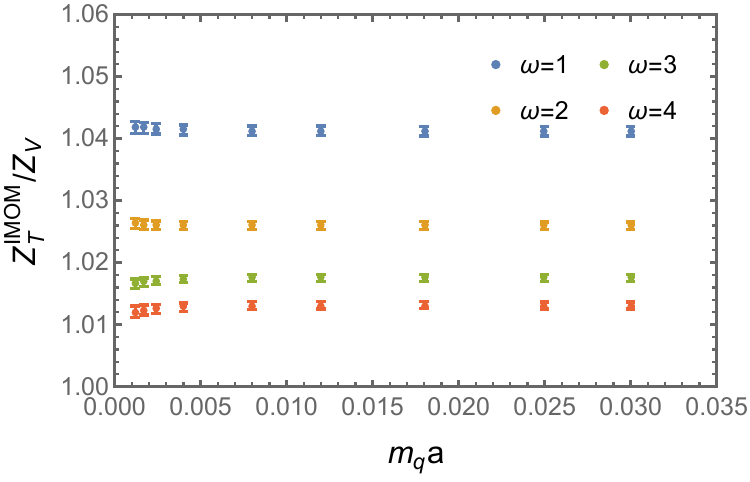}
\caption{The valence quark mass dependence of $Z_T^{\IMOM}/Z_V$ on the 64I ensemble with different $\omega$ when $a^2p^2=3.26$ for scenario A and $a^2p^2=3.13$ for scenario B.}
\label{fig:IMOM_mass_dep_t}
\end{center}
\end{figure*}

In Fig.~\ref{fig:IMOM_mass_dep}, we plot the valence quark mass dependence of $Z_S^{\IMOM}/Z_V$ on the 64I ensemble in two scenarios. Both of them show the more insensitive dependence on the quark mass compared with the $\MOM$ case. As we did in the $\SMOM$ scheme, we linearly extrapolate the results in the $\IMOM$ scheme to the chiral limit. Similarly, the tensor current case is also insensitive to quark mass as shown in Fig.~\ref{fig:IMOM_mass_dep_t}. 

The results in $\MSbar$ can be obtained by multiplying the corresponding matching factors $C_S^{\MSbar,\IMOM_\omega}$, which can be expressed as 
\begin{eqnarray}
C_{S/T}^{\MSbar,\IMOM_\omega}=1+\sum_i^3\left(\frac{\alpha_s}{4\pi}\right)^iC_{S/T,i}^{\omega},
\end{eqnarray}
and the coefficients $C_{S/T,i}^{\omega}$ are listed in Table~(\ref{tab:coeff_match_IMOM}).  The result of the $\SMOM$ case with $\omega=1$ has been calculated at the three-loop level~\cite{Kniehl:2020sgo}, while only the two-loop results are available for arbitrary $\omega$ cases~\cite{Bell:2016nar,Gracey:2019ytw,Garron:2021mfn}.

\begin{table}[htbp]
  \centering
  \vspace{0.5em}
  \begin{tabular}{c|c|c|c}
    \toprule
  $\omega$ & $C_{S,1}^{\omega}$ & $C_{S,2}^{\omega}$ & $C_{S,3}^{\omega}$ \\
\hline
1 & 0.646 & $-4.014n_f+23.024$ & $2.184n_f^2 - 169.923n_f + 889.742$ \\
2 & $-1.994$ & $1.080n_f+34.591$ & N/A   \\
3 & $-4.042$ & $1.195n_f-70.621$ & N/A\\
4 & $-5.757$ & $3.099n_f-95.751$ & N/A \\
 \hline
 \hline
  $\omega$ & $C_{T,1}^{\omega}$ & $C_{T,2}^{\omega}$ & $C_{T,3}^{\omega}$ \\
\hline
1 & $-0.215$ & $4.103n_f-43.384$ & $- 7.064n_f^2+309.829n_f-1950.761  $ \\
2 & $-0.347$ & $4.250n_f-38.902$ & N/A   \\
3 & $-0.454$ & $4.369n_f-34.277$ & N/A\\
4 & $-0.548$ & $4.464n_f-31.180$ & N/A \\
 \hline
\end{tabular}
\caption{The coefficients of the matching factors for scalar and tensor operators with different $\omega$~\cite{Bell:2016nar,Gracey:2019ytw,Kniehl:2020sgo,Garron:2021mfn}. The 3-loop coefficients with $\omega\neq 1$ are not available in the literature and marked as ``N/A'' in the table.}
  \label{tab:coeff_match_IMOM}
\end{table} 

\begin{figure*}[htbp]
	\centering
	\includegraphics[width=0.48\textwidth]{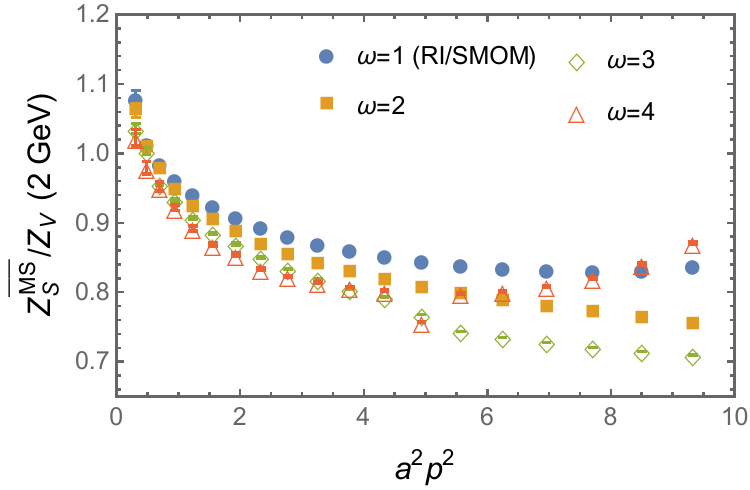}
	\includegraphics[width=0.48\textwidth]{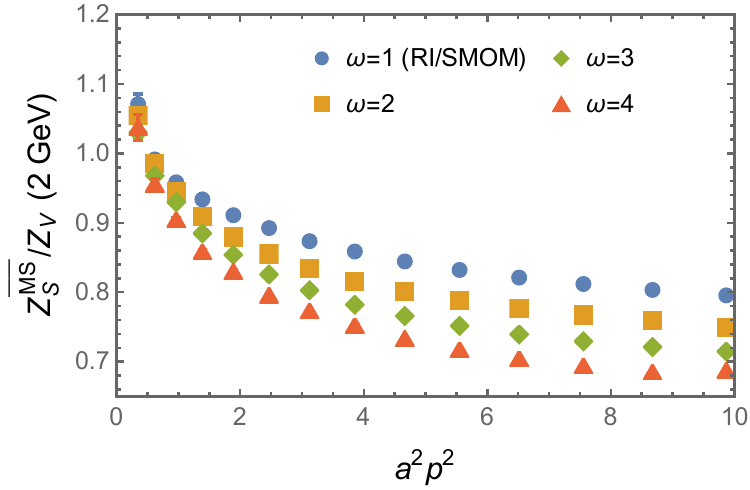}
	\caption{The results of $Z^{\MSbar}_S/Z_V$ on the 64I ensemble obtained through the intermediate schemes with different $\omega$ in different scenarios.}
	\label{fig:ZS_MSar_IMOM}
\end{figure*}

\begin{figure*}[htbp]
	\centering
	\includegraphics[width=0.48\textwidth]{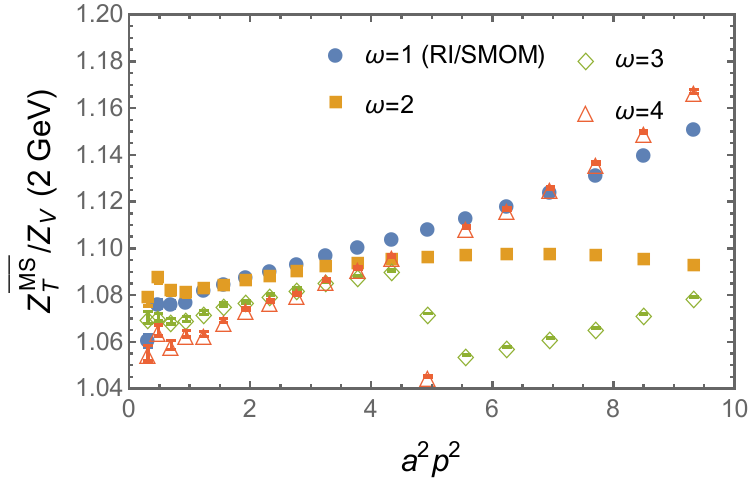}
	\includegraphics[width=0.48\textwidth]{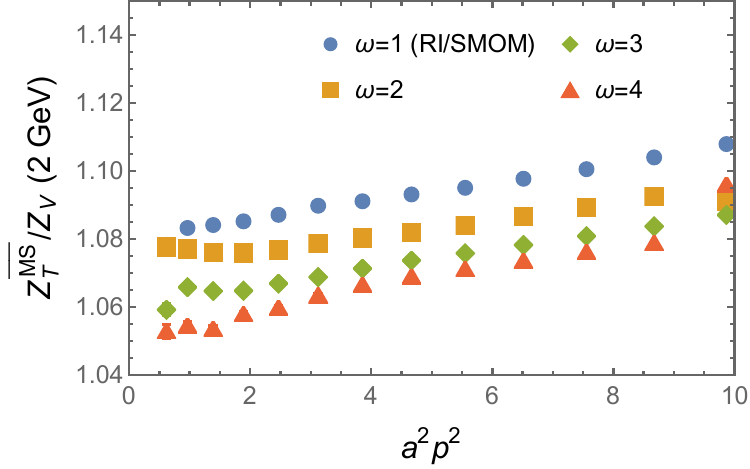}
	\caption{The results of $Z^{\MSbar}_T/Z_V$ on the 64I ensemble obtained through the intermediate schemes with different $\omega$ in different scenarios.}
	\label{fig:ZT_MSar_IMOM}
\end{figure*}

\begin{figure*}[]
\begin{center}
	\includegraphics[scale=0.65]{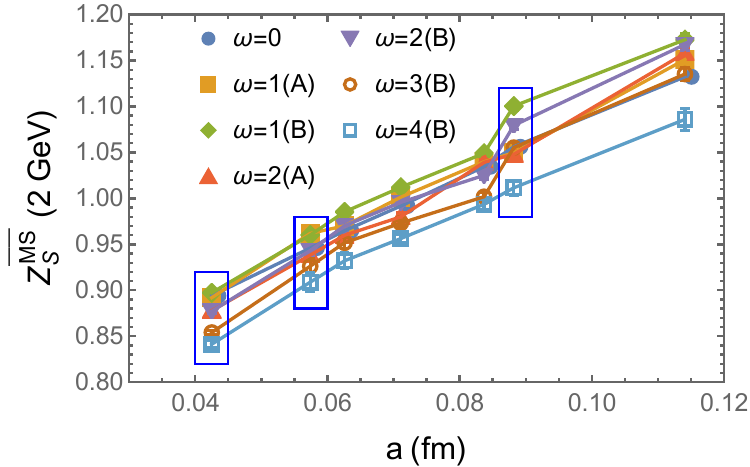}
	\includegraphics[scale=0.65]{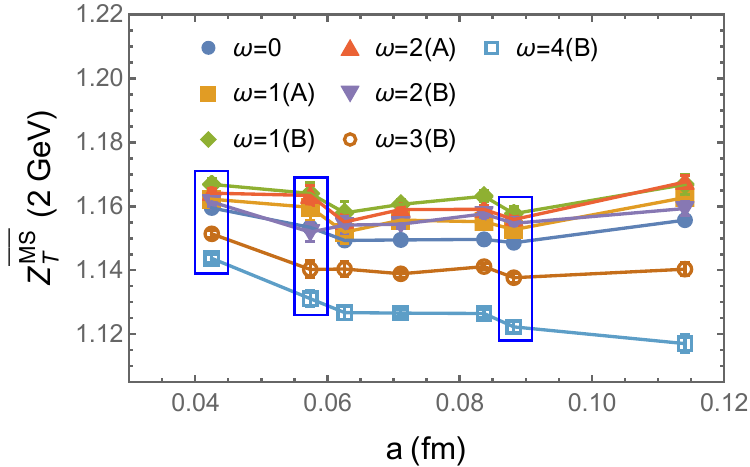}
\caption{The results of $Z^{\MSbar}_S~(2~\GeV)$ and $Z^{\MSbar}_T~(2~\GeV)$ on the different gauge ensembles. The different curves represent the results obtained from different $\omega$ choices in the $\IMOM$ schemes. The data in the blue rectangles are the results on the HISQ ensembles.}
\label{fig:IMOM_scheme}
\end{center}
\end{figure*}

Then one can evolve the results to 2 GeV using the anomalous dimension of the $\MSbar$ scheme, and obtain the $Z^{\MSbar}_S/Z_V (2~\GeV)$ shown in Fig.~\ref{fig:ZS_MSar_IMOM}, with different $\omega$ using either scenario A (left panel) or scenario B (right panel). The tensor current case is plotted in Fig.~\ref{fig:ZT_MSar_IMOM}. One can see that the cutoff effect with scenario A can introduce mutation at $a^2p^2\sim 5$ when $\omega=3$ or 4, and then it is very hard to fit the data. Using the parametrizations defined in Eq.~(\ref{eq:ZS_extrapo1}) and (\ref{eq:Zt_extrapo1}), we obtain $Z^{\MSbar}_{S,T}/Z_V (2~\GeV)$ for different $\omega$ and scenarios, and collect the results in Table~\ref{tab:sum_FR_imomZS} and ~\ref{tab:sum_FR_imomZT}, with only the statistical uncertainties. The fit ranges are same as those listed in Table~\ref{tab:sum_FR_smom} and the corresponding $\chi^2$/d.o.f.\ of fits are smaller than 1. We also illustrate the results at different lattice spacing and schemes in Fig.~\ref{fig:IMOM_scheme}, with the data on the HISQ ensembles marked with blue rectangles. As shown in the figure, the scheme dependence becomes somehow weaker at smaller lattice spacing, but not as fast as an ${\cal O}(a^2)$ effect. It means that using the IMOM scheme can be more non-trivial to control the systematic uncertainties.

\begin{table*}[htbp]
  \centering
\setlength{\tabcolsep}{0.5mm}{
  \begin{tabular}{l|c|c|c|c|c|c|c|c|c|}
  \toprule
 & \multicolumn{1}{c|}{$\omega=0$} & \multicolumn{2}{c|}{$\omega=1$} & \multicolumn{2}{c|}{$\omega=2$} & \multicolumn{1}{c|}{$\omega=3$} &\multicolumn{1}{c|}{$\omega=4$}   \\
\hline 
Ensemble & Scenario B & Scenario A & Scenario B  & Scenario A & Scenario B  & Scenario B  & Scenario B   \\
\hline 
HISQ09 & 0.975(02) & 0.969(05) & 1.016(4)  & 0.968(3)  & 0.997(4) &   0.974(6)  & 0.934(08)  \\
HISQ06 & 0.891(02) & 0.906(02) & 0.904(2) & 0.884(3) & 0.889(5) &    0.872(8)  & 0.856(10) \\
HISQ04 & 0.850(01) & 0.848(02) & 0.853(2) & 0.835(2) & 0.834(2)  &    0.812(2)   & 0.800(04) \\
48I    & 1.026(02) & 1.043(07) & 1.063(6) & 1.049(4) & 1.058(6) &    1.029(7) &   0.984(12) \\
64I    & 0.959(01) & 0.964(06)  & 0.973(3)  & 0.963(3) &  0.950(4) &    0.929(5)    & 0.921(06)  \\
48If  &  0.927(01) & 0.936(02) & 0.946(2)  & 0.916(4) & 0.931(3)  &    0.910(3)    & 0.894(05)   \\
32If  &  0.906(01) & 0.911(02) & 0.925(3)  & 0.901(6) & 0.911(5)  &    0.894(6)    & 0.875(09)  \\
\hline
  \end{tabular}}
  \caption{The fit results of $Z^{\MSbar}_S/Z_V~(2~\GeV)$ from the $\IMOM$ scheme with different $\omega$ in different scenario on the 7 gauge ensembles.}
  \label{tab:sum_FR_imomZS}
\end{table*}

\begin{table*}[htbp]
  \centering
\setlength{\tabcolsep}{0.5mm}{
  \begin{tabular}{l|c|c|c|c|c|c|c|c|c|}
  \toprule
 & \multicolumn{1}{c|}{$\omega=0$} & \multicolumn{2}{c|}{$\omega=1$} & \multicolumn{2}{c|}{$\omega=2$} & \multicolumn{1}{c|}{$\omega=3$} &\multicolumn{1}{c|}{$\omega=4$}   \\
\hline 
Ensemble & Scenario B & Scenario A & Scenario B  & Scenario A & Scenario B  & Scenario B & Scenario B   \\
\hline 
HISQ09 & 1.0604(02) & 1.0642(24)   & 1.0688(22)  & 1.0671(16)   & 1.0660(17)    & 1.0503(15) & 1.0361(17)  \\
HISQ06 & 1.0864(03) & 1.0923(41)   & 1.0964(35) & 1.0958(31) & 1.0850(29) &   1.0740(26) &    1.0654(24) \\
HISQ04 & 1.1023(01) & 1.1049(19)  & 1.1093(18) & 1.1067(13)  & 1.1041(13)    & 1.0946(11) &   1.0873(13) \\
48I    & 1.0472(04) & 1.0536(27)   & 1.0573(31) & 1.0580(21) & 1.0505(22)  & 1.0333(20)  &   1.0122(27)  \\
64I    & 1.0658(01) & 1.0709(25)   & 1.0783(17)  & 1.0745(12)  & 1.0732 (12)    & 1.0579(10)  &    1.0443(13)  \\
48If  &  1.0745(01) & 1.0803(17) & 1.0849(15)  & 1.0834(12)  & 1.0791(12)  &    1.0646(11)  &    1.0531(12)  \\
32If  &  1.0793(02) & 1.0817(34)  & 1.0874(35)  & 1.0847(23) & 1.0838(27)  &     1.0709(22) &   1.0581(21)  \\
\hline
  \end{tabular}}
  \caption{The fit results of $Z^{\MSbar}_T/Z_V~(2~\GeV)$ from the $\IMOM$ scheme with different $\omega$ in different scenario on the 7 gauge ensembles.}
  \label{tab:sum_FR_imomZT}
\end{table*}

\end{document}